\documentclass[%
 reprint,
superscriptaddress,
longbibliography,
 amsmath,amssymb,
 aps,
pra,
showkeys
]{revtex4-1}

\usepackage{graphicx}% Include figure files
\usepackage{dcolumn}% Align table columns on decimal point
\usepackage{bm}% bold math
\usepackage{hyperref}% add hypertext capabilities
\usepackage[version=4]{mhchem}
\usepackage{mathtools}
\usepackage{braket}

\DeclarePairedDelimiter\abs{\lvert}{\rvert}
\DeclarePairedDelimiter\norm{\lVert}{\rVert}
\makeatletter
\let\oldabs\abs
\def\abs{\@ifstar{\oldabs}{\oldabs*}}

\let\oldnorm\norm
\def\norm{\@ifstar{\oldnorm}{\oldnorm*}}
\makeatother

\newcommand{\const}[1]{\mathrm{#1}}

\usepackage{xcolor}

\begin{document}

\title{Sensitive single-photon test of extended quantum theory with 2D hexagonal boron nitride}

\author{Tobias Vogl}
\email{tobias.vogl@uni-jena.de}
\affiliation{Institute of Applied Physics, Abbe Center of Photonics, Friedrich-Schiller-Universit\"at Jena, 07745 Jena, Germany}
\affiliation{Cavendish Laboratory, University of Cambridge, JJ Thomson Avenue, Cambridge CB3 0HE, United Kingdom}
\author{Heiko Knopf}
\affiliation{Institute of Applied Physics, Abbe Center of Photonics, Friedrich-Schiller-Universit\"at Jena, 07745 Jena, Germany}
\affiliation{Fraunhofer-Institute for Applied Optics and Precision Engineering IOF, 07745 Jena, Germany}
\affiliation{Max Planck School of Photonics, 07745 Jena, Germany}
\author{Maximilian Weissflog}
\affiliation{Institute of Applied Physics, Abbe Center of Photonics, Friedrich-Schiller-Universit\"at Jena, 07745 Jena, Germany}
\affiliation{Max Planck School of Photonics, 07745 Jena, Germany}
\author{Ping Koy Lam}
\affiliation{Centre for Quantum Computation and Communication Technology, Department of Quantum Science, Research School of Physics and Engineering, The Australian National University, Acton ACT 2601, Australia}
\author{Falk Eilenberger}
\affiliation{Institute of Applied Physics, Abbe Center of Photonics, Friedrich-Schiller-Universit\"at Jena, 07745 Jena, Germany}
\affiliation{Fraunhofer-Institute for Applied Optics and Precision Engineering IOF, 07745 Jena, Germany}
\affiliation{Max Planck School of Photonics, 07745 Jena, Germany}


\date{\today}

\begin{abstract}
Quantum theory is the foundation of modern physics. Some of its basic principles, such as Born's rule, however, are based on postulates which require experimental testing. Any deviation from Born's rule would result in higher-order interference and can thus be tested in an experiment. Here, we report on such a test with a quantum light source based on a color center in hexagonal boron nitride (hBN) coupled to a microcavity. Our room temperature photon source features a narrow linewidth, high efficiency, high purity, and on-demand single-photon generation. With the single-photon source we can increase the interferometric sensitivity of our three-path interferometer compared to conventional laser-based light sources by fully suppressing the detector nonlinearity. We thereby obtain a tight bound on the third-order interference term of $3.96(523)\times 10^{-4}$. We also measure an interference visibility of 98.58\% for our single-photons emitted from hBN at room temperature, which provides a promising route for using the hBN platform as light source for phase-encoding schemes in quantum key distribution.
\end{abstract}

\maketitle

\section{Introduction}
\noindent Low-dimensional materials and nanostructures have gained significant attention in recent years due to the rich physics and unique properties they offer \cite{10.1088/0957-4484/19/7/075609,10.1021/acs.chemrev.6b00566,10.1021/acs.chemrev.9b00423}. Of particular interest are two-dimensional (2D) materials \cite{10.1021/cr900070d,10.1038/natrevmats.2017.33}, which can form topological insulators \cite{10.1038/s41567-018-0189-6} or Dirac semimetals \cite{10.1093/nsr/nwu080} and can exhibit properties like super-transport \cite{Sharma_2020}, strongly bound excitons at room temperature \cite{PhysRevLett.105.136805}, or superconductivity \cite{10.1038/nature26160}. This results in a broad variety of potential applications, including atomically-thin electronics \cite{10.1038/nnano.2010.279}, sensing \cite{10.1021/acsnano.6b05274,10.1002/smll.201201224}, space instrumentation \cite{10.1038/s41467-019-09219-5}, and photonics \cite{10.1021/acsami.7b09889}.\\
\indent 2D materials and their phenomena are governed by quantum physics, which revolutionized everyday life. Some of the basic principles of quantum theory, however, are based solely on axioms and postulates. One example for such fundamental principle based on a postulate is the definition of quantum mechanics on a complex Hilbert space. Another example is Born's rule: the probability density $p\left(\vec{r},t\right)$ of finding a quantum object at place $\vec{r}$ and time $t$ is given by the absolute square of its wave function $|\psi\left(\vec{r},t\right)|^2$. In Max Born's original paper this was just mentioned in a footnote \cite{10.1007/BF01397184}, yet it is now a basic foundation of quantum mechanics. The theoretical framework of quantum mechanics, however, does not require this to be true \cite{arXiv:quant-ph/0507170}. There have been efforts to derive Born's rule from more fundamental principles with an operational approach \cite{doi:10.1098/rspa.2003.1230}, using quantum logic \cite{10.2307/24900629,cooke_keane_moran_1985}, or recently only making assumptions on the mathematical structure of the measurement process in quantum mechanics \cite{Masanes2019}. Nevertheless, every approach still needs a certain level of postulation that needs to be validated.\\
\indent Any deviation from such postulates manifests in consequences that could be observed in experiments. A deviation from Born's rule leads to higher-order interference \cite{10.1142/S021773239400294X} and in a hyper-complex Hilbert space phases would not necessarily commute anymore \cite{PhysRevLett.42.683}. There have been several experimental tests of these extended quantum theories, reducing the upper bound of any potential deviation successively. Hyper-complex quantum mechanics has been tested with non-relativistic neutrons \cite{PhysRevA.29.2276} and relativistic photons \cite{10.1038/ncomms15044}. Higher-order interference has been ruled out using multi-path interferometers \cite{10.1126/science.1190545,10.1007/s10701-011-9597-5,10.1088/1367-2630/aa5d98,PhysRevResearch.2.012051,arXiv:2006.09496}. A common problem of past experiments is they require careful characterization of a complex measurement apparatus to attain significant accuracy. Noise and systematic errors in the light source or detector can lead to an apparent deviation. Moreover, the uncertainty in the characterization of the measurement apparatus always propagates into an uncertainty of the observable in question. The ultimate limits to the error are the quantum fluctuations of the light source (shot noise) and the nonlinear response of the  photodetectors \cite{10.1088/1367-2630/aa5d98}. It is possible to reduce the relative quantum fluctuations by increasing the intensity, but this also increases the nonlinearity of the detector, meaning both effects compete against each other in a classical design conflict.\\
\indent An ideal single-photon source (SPS) could resolve both issues: it is not shot noise-limited, i.e.\ it has no intensity fluctuations and consecutive single-photons can be spaced in time such that the detector response can be ensured to be linear. For a single-photon avalanche diode (SPAD), the pulse period of the photons must be chosen larger than the detector deadtime.\\
\indent In this article we use a quantum light source based on a defect center in 2D hexagonal boron nitride (hBN) coupled to a microcavity \cite{10.1021/acsphotonics.9b00314} to test for higher-order interference. The resonator improves both the spectral and photon purity, as well as ensures a high collection efficiency of the single-photons. We test for higher-order interference with a three-path interferometer with which we can determine all contributing terms to the third-order interference individually. We show that using our single-photon source we can fully circumvent the detector nonlinearity and obtain a tight bound for the higher-order interference. We also determine the intrinsic interferometer visibility of our photon source.

\section{Results}
\begin{figure}[hbt]
\includegraphics[width=\linewidth]{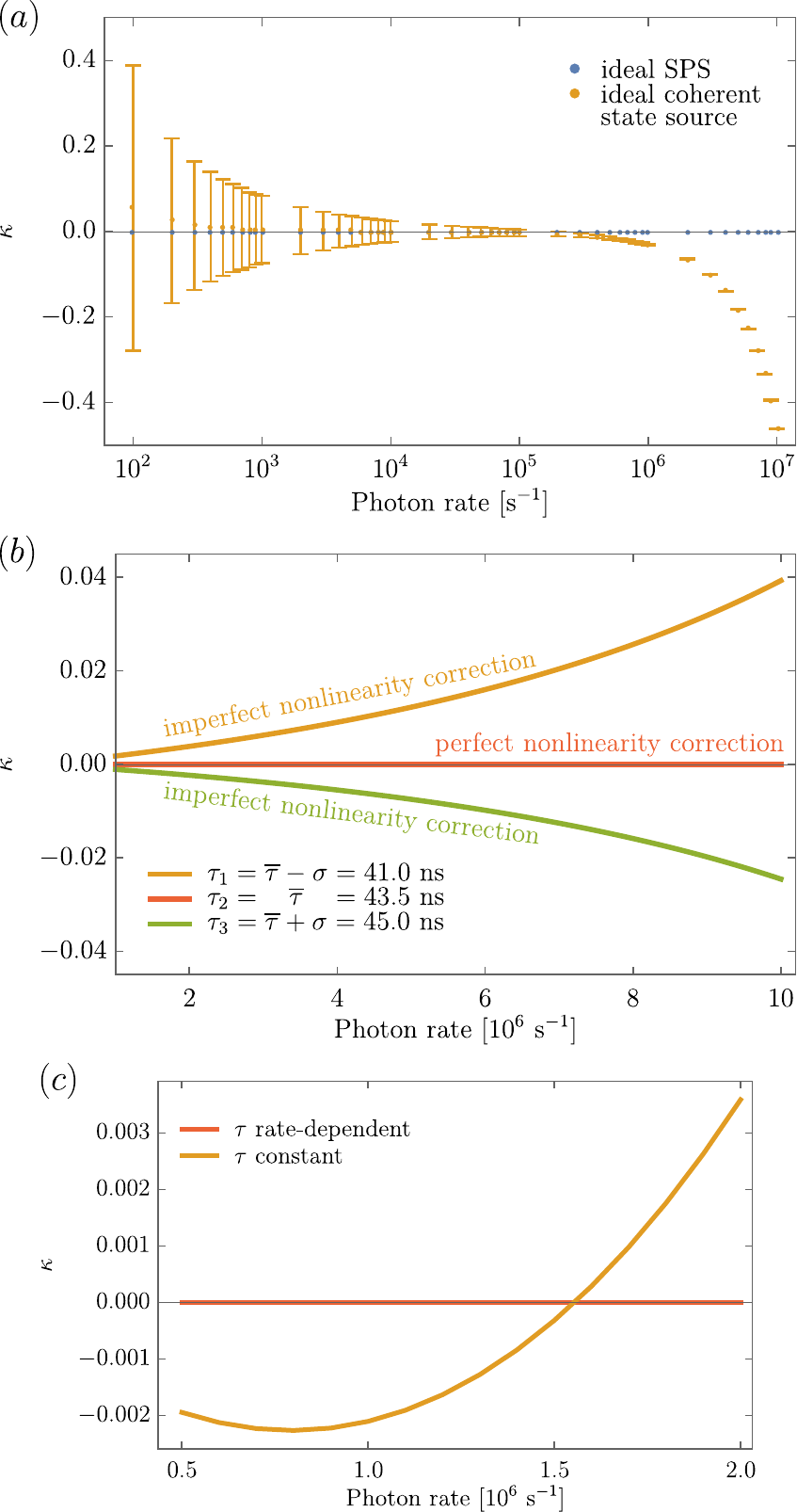}
\caption{Third-order interference $\kappa$ for realistic light sources and detectors assuming Born's rule. (a) The shot noise limit of a laser at low photon rates results in a large uncertainty and for high rates the detector nonlinearity leads to a deviation from $\kappa=0$. An ideal single-photon source is not affected by these effects. (b) Evolution of the nonlinearity-corrected $\kappa$ for different detector deadtime characterization results. The model assumes the true value to be $\overline{\tau}$ (our measured mean). (c) Evolution of $\kappa$ for a constant and a realistic rate-dependent deadtime model.} 
\label{fig1}
\end{figure}
\subsection{Theoretical treatment of quantum interference}
\noindent Quantum interference always occurs between pairs of paths (second-order interference) and there is no third-order interference. This is a direct consequence of the linearity of the underlying Hilbert space in quantum mechanics and the square-relation of Born's rule. Any deviation from Born's rule leads to higher-order interference. In Appendix \ref{appA} we follow R.\ Sorkin's theoretical treatment of quantum interference \cite{10.1142/S021773239400294X} to calculate the experimentally accessible quantity $\kappa$, which is the ratio of third-order interference to expected second-order interference. The parameter can be measured with a three-path interferometer and is given by
\begin{align*}
\kappa=\frac{\epsilon}{\delta}
\end{align*}
where the numerator
\begin{align*}
\epsilon=R_{ABC}-R_{AB}-R_{AC}-R_{BC}+R_A+R_B+R_C-R_0
\end{align*}
is the third-order interference term and the denominator
\begin{align*}
\begin{gathered}
\delta=\abs{R_{AB}-R_A-R_B+R_0}+\abs{R_{AC}-R_A-R_C+R_0}\\
+\abs{R_{BC}-R_B-R_C+R_0}
\end{gathered}
\end{align*}
is the sum of the second-order interference terms. The photon rates $R_i$ are the detected rates when the $i$th path(s) of the interferometer are open and $i=0$ corresponds to all paths closed. This parameter describes the deviation from Born's rule and allows us to compare the results from different experiments. It is worth noting the third-order interference term vanishing implies that all other higher orders vanish as well \cite{10.1142/S021773239400294X}. In order to measure $\kappa$, each path of the interferometer must be able to open and close individually.
\subsection{Realistic measurement apparatus}
\noindent As already mentioned, an imperfect measurement apparatus can lead to apparent higher-order interference. In Fig.\ \ref{fig1}(a) we simulate how $\kappa$ evolves as a function of the incident photon rate into the three-path interferometer for an ideal coherent state source (laser) and an ideal SPS. To simulate a (shot noise-limited) coherent state source, we perform Monte Carlo simulations and average over a sample size of $N=10^4$ (for each rate). To model the nonlinearity, a detector deadtime of 45 ns has been assumed (see Appendix \ref{appB} for the nonlinear model). The sample size $N$ was chosen such that the mean value of $\kappa$ does not significantly change when increasing the sample size to $N+1$:
\begin{align*}
\frac{1}{N}\sum_i^N \kappa_i - \frac{1}{N+1}\sum_i^{N+1} \kappa_i \ll 1
\end{align*}
For low photon rates shot noise dominates and the uncertainty of $\kappa$ (standard deviation over all simulation runs) is large, thus providing only a weak upper bound on the third-order interference. When averaging over a large sample, the mean value can be estimated better, the large uncertainty, however, will remain. It should be noted that the bias of $\kappa$ in the low photon rate limit is a systematic error also caused by shot noise (see Appendix \ref{appMC} for the full explanation). As the photon rate increases, the relative intensity fluctuations decrease and consequently also the uncertainty of $\kappa$, but eventually the detector nonlinearity leads to an apparent and systematic deviation from $\kappa=0$. With an ideal SPS one would find $\kappa=0$ for all photon rates. While shot noise is not present in an ideal SPS, the detector deadtime can contribute towards a finite $\kappa$. Nevertheless, by adjusting the time interval of consecutive single-photons, this effect can be prevented. It is worth noting that the data in Fig.\ \ref{fig1}(a) for both light sources has not been corrected for the detector nonlinearity. While a single-photon source does intrinsically not require such correction, for the coherent state source this is attempted in the following.\\
\indent To account for this systematic error for the coherent state source, the deadtime must be known, however, this requires an accurate characterization. Furthermore, it is important that this measurement does not rely on interference, otherwise any higher-order interference term would already influence this characterization. We have characterized our SPADs (ID Quantique ID100) with the superposition method \cite{10.1063/1.4879820} (see Appendix \ref{appB}) and found a deadtime of $\tau=43.56(143)$ ns. We have used the standard deviation of multiple measurement repetitions as the uncertainty. The question arises how such uncertainty propagates into $\kappa$. For the following model we assume for simplicity that our measured mean value of 43.5 ns is the true detector deadtime. In general this will not be the case, since approximating the mean exactly takes infinite measurements. We then calculate the nonlinearity-corrected $\kappa$ for the potential characterization results $\tau_1=41.0$, $\tau_2=43.5$, and $\tau_3=45.0$ ns, corresponding to our measured mean $\pm$ one standard deviation, respectively. As shown in Fig.\ \ref{fig1}(b), even after the correction, the uncertainty leads again to an apparent and systematic third-order interference strength on the order of 1\% of the second-order interference for photon countrates exceeding $10^6$ s$^{-1}$. This is possibly a limiting factor in a previous experiment \cite{10.1007/s10701-011-9597-5}.\\
\indent In addition to the potentially inaccurate characterization, the detector deadtime is countrate-dependent and becomes shorter near saturation. This is due to a charge/discharge of a fixed RC circuit and a comparator with a set threshold in the detector. When the countrate increases, the RC circuit cannot fully charge/discharge, resulting in a slight decrease of the deadtime. This makes an accurate characterization impractical. To illustrate this countrate-dependency of the deadtime, we have repeated the characterization of the SPADs with the superposition method for different photon countrates. Of course, each measurement for each photon countrate has now its own uncertainty, propagating into an uncertainty of our extracted rate-dependent model. For the following we again assume our model describes the true nonlinear behavior of the detector, but in general this is not the case. When correcting $\kappa$ with the rate-dependent deadtime (for the rate-dependent deadtime model see Appendix \ref{appB}) there is of course no deviation (see Fig.\ \ref{fig1}(c)), but when we correct with a constant deadtime, this leads to a value for $\kappa$ on the order of $10^{-3}$. This is, in fact, also the current state-of-the-art in the quantum regime \cite{10.1088/1367-2630/aa5d98}. It should be noted that even the  correction of the rate-dependent deadtime with a constant deadtime yields $\kappa=0$ for a rate of 1.55 MHz. This is due to the fact that the constant deadtime is the correct deadtime for one particular rate and hence there is a combination of rates which yields $\kappa=0$. We note that the input rate at which this happens cannot be estimated accurately, as this requires accurate characterization of the deadtime and if this was possible, accurate deadtime correction could be employed in the first place. Furthermore, the detector deadtime characterization using the superposition method is also wavelength dependent, making an accurate characterization even more difficult (see Appendix \ref{appB}).\\
\indent Using a true single-photon source or quantum emitter, these problems can be circumvented completely. For a sufficiently low second-order correlation function $g^{(2)}\left(0\right)$ and high enough quantum efficiency, the quantum light source beats the shot noise limit and each single-photon pulse contains one and only one photon, i.e.\ there are no fluctuations. Moreover, spacing the photons longer in time than the detector deadtime makes the detector response perfectly linear. The latter can be easily achieved by pulsed single-photon generation and choosing a pulse period longer than the deadtime. For our detector this limits the maximal incident rate to 20 to 25 MHz. This of course implies that preventing the deadtime only works up to a maximal photon rate, however, this maximum is anyway the detection limit of the specific detector due to the deadtime. In other words, with the single-photon source the nonlinearity can be prevented over the entire dynamic range of the detector. Alternatively, when the excited state lifetime of the quantum emitter is much longer than the deadtime, the single-photon source could also be excited continuously.

\subsection{Quantum light source}
\begin{figure*}[hbt]
\includegraphics[width=\linewidth]{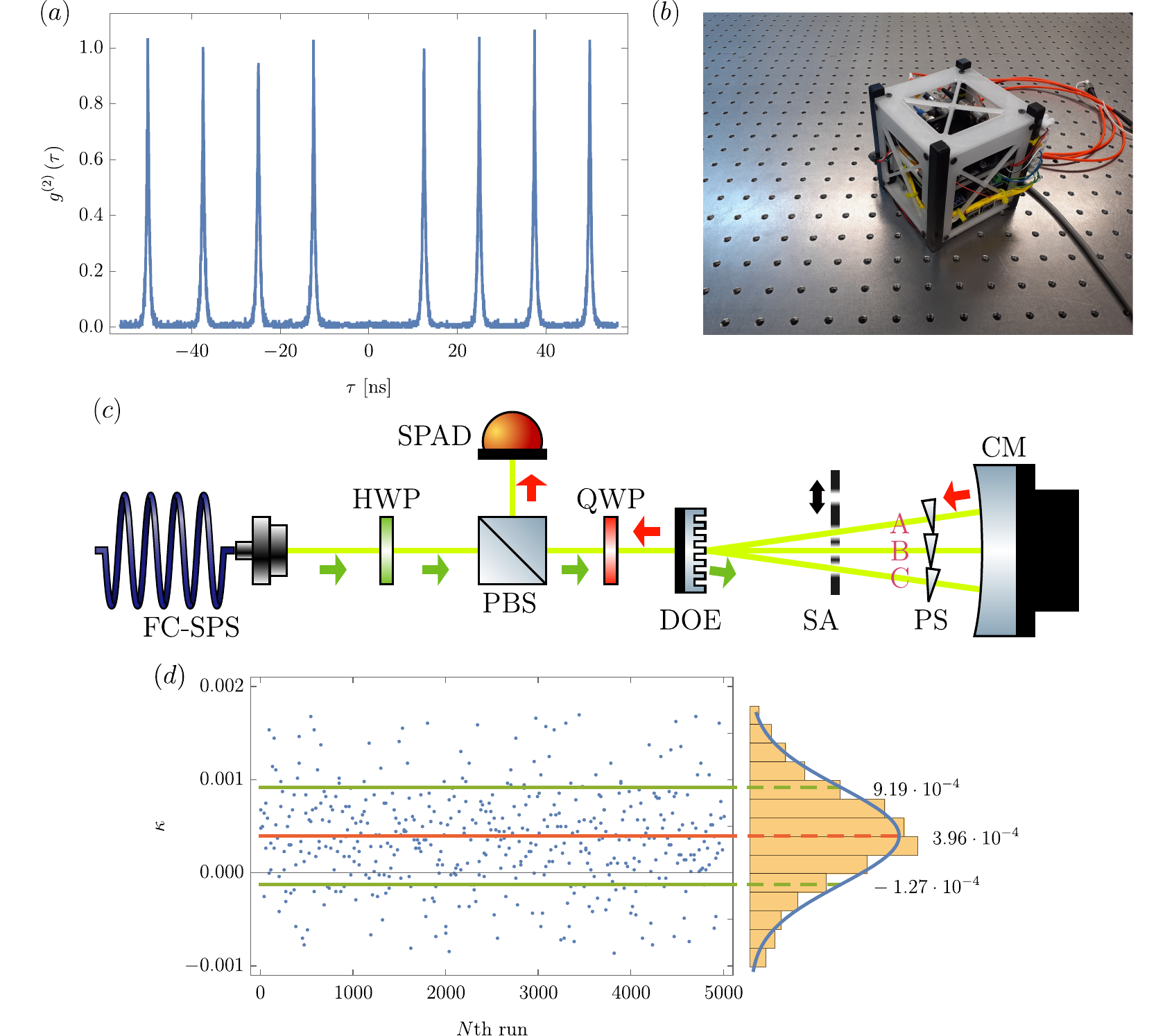}
\caption{Experimental setup and results. (a) Second-order correlation function recorded with pulsed excitation. The data has been centered around zero time delay and normalized. Integration over the peaks and the same area around zero yields $g^{(2)}\left(0\right)=6.4\times 10^{-3}$ after background correction. (b) Photograph of the integrated quantum light source mounted in a 1U CubeSat structure. (c) The polarization of the photons from the fiber-coupled single-photon source (FC-SPS) is rotated with half-wave plate (HWP) and additionally filtered by a polarizing beamsplitter (PBS). A quarter-wave plate (QWP) rotates the polarization to left-circular. A diffractive optical element (DOE) splits the photon amplitude into three equal ones. A concave mirror (CM) retroreflects the photons which recombine at the DOE and exit the PBS on the reflective port (due to the phase shift of the reflection at the CM), where the photons are detected by a SPAD. The forward/reflected propagation direction is marked with green/red arrows. A shutter array (SA) can be moved through the paths, allowing us to open/close each path (A, B, C) individually. The phase shifters (PS) are optional for the determination of the interference visibility. (d) Measurement repetitions of $\kappa$. The mean value of 5000 runs is $3.96\times 10^{-4}$ (red line) and the standard deviation is $5.23\times 10^{-4}$ (green lines). These values have been determined from fitting a normal distribution to the histogram (on the right). For the sake of clarity, the data shown here has been downsampled by a factor of 11.}
\label{fig2}
\end{figure*}

\noindent Our quantum light source is based on a color center in hBN, which is an intensively researched single-photon emitter system due to its superior brightness \cite{nnano.2015.242,10.1021/acsnano.6b03602,10.1088/1361-6463/aa7839}, purity \cite{10.1021/acsphotonics.8b00127}, and quantum efficiency \cite{10.1364/OPTICA.6.001084} at room temperature. In addition, the emitters can have near-ideal out-coupling efficiency, with no photons lost due to total internal or Fresnel reflection \cite{C9NR04269E}. Recent results suggest the quantum emission in the visible spectrum is carbon-related \cite{arXiv:2003.00949}. Long-time characterizations proved their reliability \cite{10.1021/acsphotonics.8b00127,LIU2020114251}, allowing us to use the emitter for the extended quantum theory test. While there has been a lot of research into the creation \cite{10.1021/acs.nanolett.6b03268,10.1021/acsami.6b09875}, characterization \cite{PhysRevB.98.081414}, and tuning of the emitters \cite{10.1021/acs.nanolett.8b01030,10.1021/acsnano.8b08511}, applications are still sparse. So far, there has been quantum random number generation \cite{arXiv:2001.10625} and nanothermometry \cite{10.1021/acsami.0c05735} based on single-photon emitting hBN.\\
\indent To increase the spectral and photon purity, as well as the collection efficiency, we have coupled our emitter to a plano-concave microcavity (for details see Ref.\ \cite{10.1021/acsphotonics.9b00314}). The defect is excited off-resonantly outside the stopband of the cavity with an ultrashort pulse laser at 470 nm where the cavity is transparent for the excitation. We have used the second-harmonic of a tunable Ti:sapphire laser at a repetition rate of 80.136 MHz. We estimate the pulse length of the second-harmonic to be well below 200 fs. This ultrashort pulse length guarantees that there is no double-excitation during a single laser pulse (the probability for a double-excitation is $5\times 10^{-4}$ at full saturation). The single-photon linewidth is 224 pm (FWHM) at a center wavelength of 565 nm. The center wavelength can be tuned by adjusting the cavity length within the free-space emission linewidth of the emitter, which is 5.76 nm (FWHM). The second-order correlation function as shown in Fig.\ \ref{fig2}(a) reveals a very high photon purity with $g^{(2)}\left(0\right)=6.4\times 10^{-3}$ after background correction and in the linear power limit. The single-photons are linearly polarized and are separated from the excitation laser by a dichroic mirror and an additional band-pass filter. The single-photon source provides two fiber output ports after a beamsplitter, one for a locking loop to keep the cavity resonant with the emitter and the other guiding the photons to the actual experiment into the interferometer. The entire light source is fully self-contained within a volume less than 1 liter and weighs 595 g (including housing, all optics, electronics, and an onboard computer for tuning of the source). This does not include the ultrashort pulse laser and second-harmonic stage used for excitation, however, the light source is also equipped with a CW laser which is included within the volume. A photograph of the integrated light source can be found in Fig.\ref{fig2}(b). The excitation laser can be switched by connecting a single-mode fiber.

\subsection{Testing for higher-order interference}
\noindent We now turn to the measurement of $\kappa$. The experimental setup for the higher-order interference test is shown in Fig.\ \ref{fig2}(c): the single-photons from our quantum light source are guided by a fiber to the interferometer. While the single-photons as emitted by the cavity are linearly polarized, the fiber rotates this polarization by a random but fixed angle. Using a half-wave plate and a polarizing beamsplitter (PBS), this polarization is reset to linear vertical polarization and then with a quarter-wave plate (QWP) angled at -45$^\circ$ to left-circular polarization. The photon wave is split into three amplitudes of equal strength by a $1\times 3$ diffractive grating beamsplitter (see Appendix \ref{appC}). The light in each path is retroreflected by a concave silver mirror ($R=100$ mm) and then recombines back at the grating. Using a concave mirror ensures equal path lengths by default and hence also operation in an interference maximum. This in turn reduces the influence of phase instability in the measurements. As the phase shift induced by the mirror exchanges left- and right-handed circular polarization, the reflected photons are horizontally polarized after the QWP and thus exit the initial PBS on the reflective port. Finally, a single-photon avalanche diode detects the photons. The initial alignment of the setup was done using a 561 nm laser system with a coherence length exceeding 100 m.\\
\indent The separation angle after the grating is $8.08^\circ$ and a mask moved through the beams allows us to block each path individually. To gather large statistics, we have mounted the blocking mask on a fast motorized stage (Thorlabs DDS220) and automated the entire experiment. The large separation angle ensures a macroscopic distance of the individual paths in the interferometer (12.5 mm separation at the blocking mask), thus guaranteeing independence of the paths and preventing any potential crosstalk or nonclassical paths \cite{PhysRevLett.113.120406,PhysRevA.85.012101}.\\
\indent We first measured the single-photon interference visibility. Depending on the setting of the mask position, we can emulate three different two-path Michelson interferometers and one three-path Michelson-type interferometer. To vary the phases in the interferometer arms, we have used phase plates with a custom-made anti-reflection coating on both sides (see Appendix \ref{appD}). All phase plates have been coated within the same coating run to minimize differences between individual phase plates. Due to the slightly unbalanced splitting ratios of the grating (see Appendix \ref{appC}), the interference visibilities are limited, but still theoretically all exceed 99.93\% (see Tab.\ \ref{tab:visibility}). Thus, the interference visibility can be used to characterize the coherence of the single-photon source. For details of the theoretical calculations see Appendix \ref{appE}.\\
\indent The experimental characterization shows high interference visibilities up to 98.58\% (see Tab.\ \ref{tab:visibility}). To the best of our knowledge, this is the first time that such interference visibility has been measured for single-photons emitted from hBN at room temperature. We do note, however, there was a measurement of this interference visibility for single-photons from cryogenically cooled quantum emitters at a temperature of 10 K \cite{PhysRevResearch.2.012016}.\\
\indent In the final experiment we measured $\kappa$ directly. To ensure that the interferometer is most sensitive to any potential higher-order interference, the angles of the phase plates were set to the central interference maximum. Due to the fixed repetition frequency of the ultrashort pulse laser, for this experiment we replace the excitation with another laser system with a variable repetition frequency ranging from 2.5 to 80 MHz. To prevent more than one pulse during the detector deadtime we choose 10 MHz as the excitation frequency. As already mentioned, the automatization of the experiment allows us to repeat the experiment many times and record large data sets that we can average over. The order in which the individual paths were opened and closed has been randomized for each run to reduce potential systematic errors. We have used data acquisition times of $1,2,5$ s per configuration without noticing any significant difference. One specific measurement campaign is shown in Fig.\ \ref{fig2}(d), with $\kappa = 3.96(523)\times 10^{-4}$, averaged over 5000 runs (corresponding to a total measurement time of 58 hours). The mean and standard deviation have been obtained by fitting a normal distribution to the histogram of the measured $\kappa$ values (see Fig.\ \ref{fig2}(d) on the right). The mean is marked in red and mean $\pm$ one standard deviation is marked in green.
\begin{table}[tbp]
\centering
\caption{Theoretical (based on the grating splitting ratios) and experimentally measured interferometric visibilities. The individual paths A, B, and C are labeled in Fig.\ \ref{fig2}(c).}
\label{tab:visibility}
\begin{ruledtabular}
\begin{tabular}{ccc}
%\hline
Paths &  Theory [\%] & Experiment [\%] \\
\colrule
AB & 99.9429 & 98.56\\
AC & 99.9369 & 95.17\\
BC & 99.9998 & 98.58\\
ABC & 99.9987 & 98.22\\
%\hline
\end{tabular}
\end{ruledtabular}
\end{table}

\section{Conclusion and outlook}
\noindent We have tested a fundamental principle of standard quantum theory based on a postulate experimentally. In particular, we have tested for deviations from Born's rule, which would manifest in higher-order interference. We demonstrated that using a single-photon source based on 2D hexagonal boron nitride coupled to a resonator, we can effectively circumvent the shot noise limit and the detector nonlinearity, which have been the limiting effects in past experiments. Thus, we have been able to get an accurate upper bound to the amount of higher-order interference, which was $3.96(523)\times 10^{-4}$ compared to the expected second-order interference. For the third-order interference term in the quantum regime, this upper bound is tighter than the previous best value of $-1.1(16)\times 10^{-3}$ \cite{10.1088/1367-2630/aa5d98}. Our experiment is also the first experiment of this kind utilizing a true single-photon source (excluding heralded spontaneous parametric down-conversion \cite{10.1126/science.1190545,10.1088/1367-2630/aa5d98}).\\
\indent The quantum light source used in the experiment was already compact and fully self-contained within a volume of less than 1 liter and a weight of less than 1 kg. The light source therefore fulfills the strict requirement for a 1U CubeSat, a picoclass satellite standard. It is hence possible to miniaturize the interferometer and test for higher-order interference in microgravity on a CubeSat in space. Potential gravitational effects could thus be ruled out as well.\\
\indent The quantum light source could also be employed to other tests of quantum theories beyond the standard model. The same detector nonlinearity effect is skewing the interference contrast of Mach-Zehnder and Sagnac interferometers in photonic tests of hyper-complex quantum mechanics. The interference pattern of these interferometers is of the form of a sine squared (see Appendix \ref{appE}) and thus requires a high dynamic range of the detectors. Circumventing the detector nonlinearity and reducing intensity fluctuations could improve the accuracy of such measurements. This also demonstrates a quantum advantage of using single-photons for interferometric sensing applications.\\
\indent Our results also suggest, that the single-photons can be used for phase-encoding schemes for quantum key distribution (QKD). The quantum bit error ratio (QBER), the defining metric in QKD, is directly related to the interference visibility $V$ for phase encoding \cite{RevModPhys.74.145}:
\begin{align*}
\text{QBER}=\frac{1-V}{2}
\end{align*}
The intrinsic limit on the QBER for our photon source is thus 0.709\%. This is important as it has been proven recently that single-photons from hBN can outperform state-of-the-art weak coherent and decoy protocols for QKD \cite{10.1021/acsphotonics.9b00314}. This extends the possible applications of single-photon sources based on hBN and offers a promising route to versatile optical quantum technologies with the hBN platform.

\section*{Acknowledgments}
\noindent We thank L\'ea Carmagnolle and the ID Quantique team for fruitful discussions on characteristics of single-photon avalanche diodes. This work was funded by the Australian Research Council (CE170100012, FL150100019), the Federal Ministry of Education and Research BMBF (13XP5053A), and by the European Union, the European Social Funds and the Federal State of Thuringia (2019FGR0101).

%\clearpage
%\onecolumngrid
%\renewcommand\thesection{S\arabic{section}}
%\setcounter{section}{0}
%\renewcommand\thetable{A\arabic{table}}
%\setcounter{table}{0}
%\renewcommand\thefigure{A\arabic{figure}}
%\setcounter{figure}{0}
%\pagenumbering{arabic}
%\renewcommand*{\thepage}{S\arabic{page}}
%\section*{Supplementary Material}
%\clearpage
\appendix

\section{Theoretical treatment of quantum interference}\label{appA}
\noindent The calculation follows the theoretical treatment of quantum interference by R.\ Sorkin \cite{10.1142/S021773239400294X}, which was also used in past experiments \cite{10.1126/science.1190545,10.1007/s10701-011-9597-5,10.1088/1367-2630/aa5d98}. According to Born's rule the probability density after $k$ slits is given by
\begin{align}
p=\abs{\sum^k_i \psi_i}^2
\end{align}
where we have dropped the spatial and temporal dependence for the sake of conciseness. Thus, Born's rule predicts quantum interference occurs only between pairs of paths and there is no higher-order interference. For a double-slit experiment with slits $A$ and $B$ we find
\begin{align}
p_{AB}=\abs{\psi_A+\psi_B}^2=\abs{\psi_A}^2+\abs{\psi_B}^2+\psi_A^*\psi_B+\psi_A\psi_B^*
\end{align}
The complex cross terms can be defined as the second-order interference term:
\begin{align}
I_{AB}&=\psi_A^*\psi_B+\psi_A\psi_B^*=p_{AB}-\abs{\psi_A}^2-\abs{\psi_B}^2=\nonumber\\
&=p_{AB}-p_{A}-p_{B}
\end{align}
Adding more paths to the interferometer, meaning opening more slits, does not increase the complexity. For example, for three slits the probability distribution at the detection screen is given by
\begin{align}
p_{ABC}=p_A+p_B+p_C+I_{AB}+I_{AC}+I_{BC}
\end{align}
In analogy to the second-order interference term, we can define the third-order term as
\begin{align}
I_{ABC}&=p_{ABC}-(p_A+p_B+p_C+I_{AB}+I_{AC}+I_{BC})=\nonumber\\
&=p_{ABC}-p_{AB}-p_{AC}-p_{BC}+p_A+p_B+p_C
\end{align}
Hence, by measuring each contributing term individually it is possible to measure $I_{ABC}$. In standard quantum theory, we find of course $I_{ABC}=0$. It is worth noting, if the third-order interference term vanishes, all other higher orders vanish as well \cite{10.1142/S021773239400294X}. In a practical experiment, the probabilities are not directly accessible. It is convenient to measure photon countrates $R_i$ (where the $i$th slits are open) instead. These rates are directly proportional to the probabilities, which yields the parameter
\begin{align}
\epsilon=R_{ABC}-R_{AB}-R_{AC}-R_{BC}+R_A+R_B+R_C-R_0
\end{align}
Note that each of the seven terms has been corrected for the darkcount rate, corresponding to all slits being closed. The parameter $\epsilon$ alone is not very meaningful, as it depends on the incident photon countrate, which is different for every individual experiment. To compare the parameter to other experiments, it is common to normalize this third-order interference by the sum of the expected second-order interference
\begin{align}
\begin{gathered}
\delta=\abs{R_{AB}-R_A-R_B+R_0}+\abs{R_{AC}-R_A-R_C+R_0}\\
+\abs{R_{BC}-R_B-R_C+R_0}
\end{gathered}
\end{align}
This gives the parameter 
\begin{align}
\kappa=\frac{\epsilon}{\delta}\label{eqn8}
\end{align}
that is directly accessible through an experiment. All that is required is a three-path interferometer where each path can be opened and closed individually. In this way, each of the eight contributing terms to $\kappa$ in Eqn.\ \ref{eqn8} can be measured.

\section{Characterization of the detector deadtime}\label{appB}
\noindent The single-photon avalanche diodes have a nonlinear response function of the form
\begin{align}
R_{\text{det}}=\frac{R_{\text{act}}}{1+\tau R_{\text{act}}}+R_0\label{eqn9}
\end{align}
where $R_{\text{det}}$ is the detected photon countrate, $R_{\text{act}}$ is the actual incident photon rate, $\tau$ is the detector deadtime, and $R_0$ is the dark countrate. For the sake of clarity we have assumed the detector efficiency to be 100\%. The detector deadtime has been characterized using the superposition method \cite{10.1063/1.4879820}. It is particularly important that this characterization is not relying on an interference effect. Otherwise, any higher-order interference terms would already influence the deadtime measurement. In the superposition method two lasers with different wavelengths to prevent interference are combined with a beamsplitter. We have used a 532 and 561 nm commercial laser system, which have been intensity stabilized. After the beamsplitter, the light is focused onto the detector. With a motorized blocking mask prior to the beam combination we can block each laser individually. To measure the deadtime, all four possible combinations of blocked/unblocked lasers have to be measured. A linear detector would satisfy
\begin{align}
R_{\text{det}}^{532+561}=R_{\text{det}}^{532}+R_{\text{det}}^{561}
\end{align}
A measurement with a realistic detector, however, will deviate from this by some residual $r$:
\begin{align}
R_{\text{det}}^{532+561}-R_{\text{det}}^{532}-R_{\text{det}}^{561}=r
\end{align}
To gather sufficient statistics, we have repeated each measurement 100 times. A simple direct optimization of the minimization condition
\begin{align}
\min \left(\sum_i^N r_i^2\right)
\end{align}
allows us to extract the detector deadtime using Eqn.\ \ref{eqn9}. We have also varied the laser powers to measure the deadtime at different photon countrates. The deadtime $\tau$ decreases with increasing countrate $R$ (see main text):
\begin{align}
\tau\left(R\right)=51.8\text{ ns}-3.333\text{ fs}\times R
\end{align}
\noindent It should be noted that this model is not valid anymore near saturation, where the dependency of the deadtime on the countrate becomes sublinear.\\
\indent It is worth noting that the detection efficiency (which we have set to 100\% for the calculations) in general is wavelength-dependent, leading to an error when using different wavelengths for the deadtime characterization. The deadtime is higher at wavelengths where the detector is more efficient: a photon that is not detected, i.e.\ a photon that does not trigger an avalanche is also not switching the detector into its dead-state. According to the manufacturer, the detection efficiency at 532 nm is 34\%, while at 561 nm it is 32\%. The overall error we make by assuming a constant efficiency is therefore rather small.

\section{Monte Carlo simulations of an ideal coherent state source}\label{appMC}
\begin{figure}[!htb]
\includegraphics[width=\linewidth]{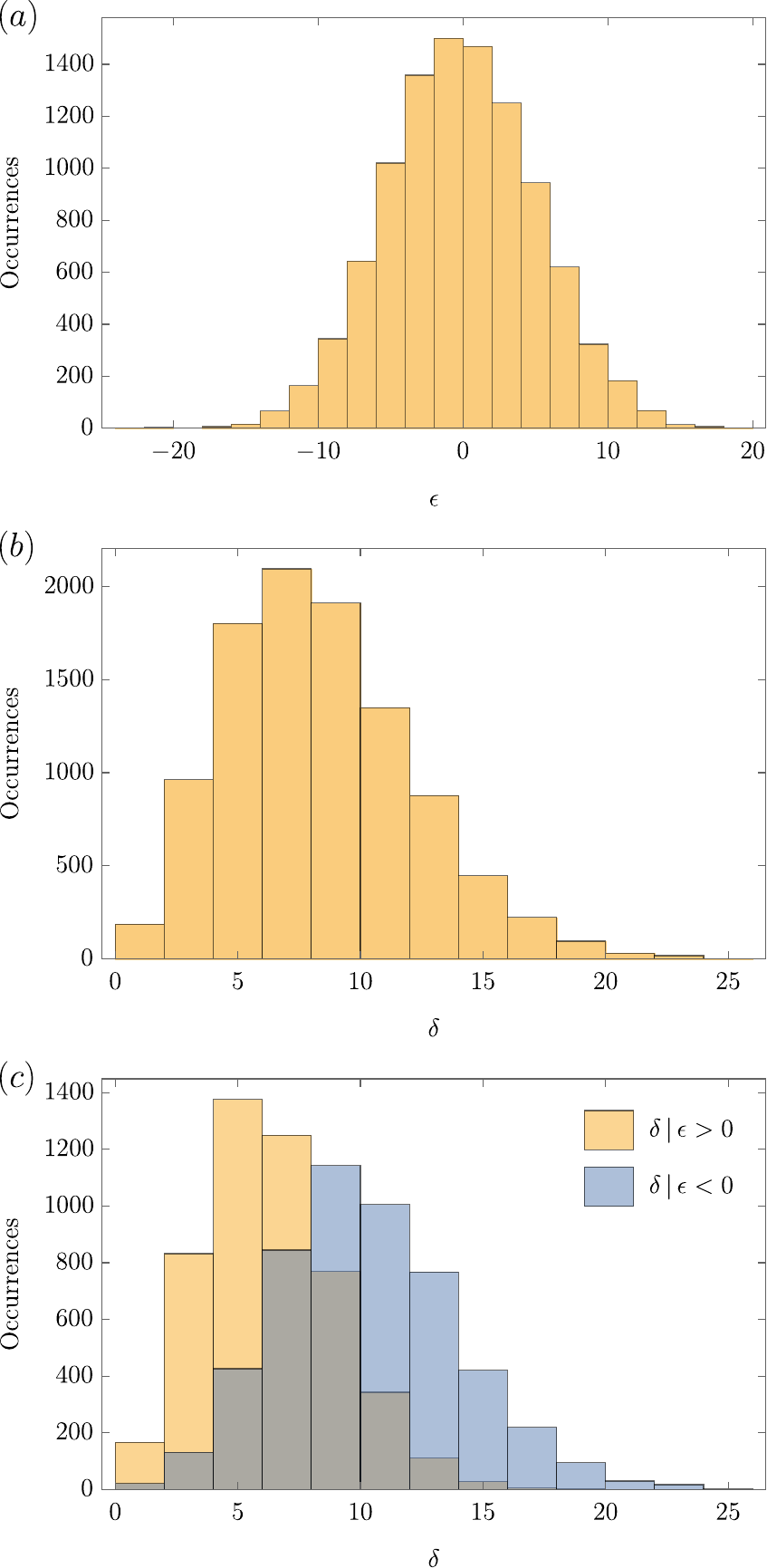}
\caption{Histograms of simulated parameters $\epsilon$ (a) and $\delta$ (b) for a shot-noise-limited light source at a photon rate of 10 Hz. Even though these parameters are normally distributed, they are not independent from each other: the mean value of $\delta$ for the condition $\epsilon>0$ is smaller (6.345) compared to the mean value of $\delta$ for the condition $\epsilon<0$ (10.28), as can be seen by the yellow and blue histograms in (c), respectively. This in turn leads to a bias in $\kappa$ when averaging over all simulations.} 
\label{figA1}
\end{figure}

\noindent It is worth noting that the Monte Carlo simulation (see main text) of the parameter $\kappa$ (averaged over the entire sample size) for the coherent state source predicts a deviation from $\kappa=0$ at low photon rates. This is not a numeric artifact caused by an insufficiently large sample size, but rather an actual systematic deviation that appears for coherent states. Since all $R_i$ are drawn from a normal distribution one might rightfully assume that $\epsilon$ (see Fig.\ \ref{figA1}(a)) and $\delta$ (see Fig.\ \ref{figA1}(b)) are also normally distributed. The parameters $\epsilon$ and $\delta$ are not independent from each other: $\epsilon<0$ is when $\abs{-R_{AB}-R_{AC}-R_{BC}}$ is large, however, this implicates that $\delta$ is large as well. In other words, a small (particularly a negative) $\epsilon$ gets divided by a large $\delta$, leading to a positive bias of $\kappa$. This is visualized by the distribution of $\delta$ for the condition $\epsilon<0$ and $\epsilon>0$ (see Fig.\ \ref{figA1}(c)). For this particular simulation presented here, the mean value of all $\epsilon$ is $\overline{\epsilon}=-0.09222$ and the mean value of all $\delta$ is $\overline{\delta}=8.361$. However, the mean value of all $\delta$ when $\epsilon<0$ is $\overline{\delta}\,|\,\epsilon<0=10.28$, while the mean value of all $\delta$ when $\epsilon>0$ is $\overline{\delta}\,|\,\epsilon>0=6.345$. Overall, this means that for a coherent state source at low incident photon rates there is not only a large uncertainty of $\kappa$, but also a systematic error, both due to shot noise. A single-photon source is not shot noise-limited, hence there is neither an uncertainty of $\kappa$, nor is there any systematic deviation from $\kappa=0$ (predicted by standard quantum theory).

\section{Characterization of the grating}\label{appC}
\noindent A custom-made diffractive optical beamsplitter (TS-210-562-Y-A) manufactured by HOLO/OR has been used. The grating is designed to split a monochromatic beam into three beams without changing any of the incident light characteristics (such as polarization), except for its power and propagation direction. The separation angle is 8.08$^\circ$ and the grating has an anti-reflective coating for 562 nm on both sides. We have characterized the grating transmission efficiency of all orders and compare it to theoretical design values (see Tab.\ \ref{tab:grating}): according to the specification provided by the manufacturer, the efficiency into the 0th order is 28.65\% and 28.35\% into the $\pm 1$st order. Experimentally, we found 29.28\%, 28.28\%, and 29.30\% for the 0th and $\pm$ 1st orders, respectively. We note that the overall transmission is nearly 100\%, i.e.\ the grating is lossless (within the experimental accuracy). However, there the amplitudes of higher orders are non-zero. A polarization-resolved characterization confirmed the polarization insensitivity within about 0.1\%. The data presented in Tab.\ \ref{tab:grating} was measured with left-circular polarization, the same polarization that was used in the experiments.

\begin{table}[htbp]
\centering
\caption{Grating efficiency and uniformity. The overall transmission is close to 100\%, but there is a finite transmission into higher orders. The uniformity is the standard deviation of the diffraction orders.}
\label{tab:grating}
\begin{ruledtabular}
\begin{tabular}{ccc}
%\hline
Order &  Theory [\%] & Experiment [\%] \\
\colrule
-1 & 28.35 & 28.28\\
0 & 28.53 & 29.26\\
+1 & 28.35 & 29.30\\
Overall & 85.23 & 86.84\\
Uniformity & 0.10 & 0.58\\
%\hline
\end{tabular}
\end{ruledtabular}
\end{table}

\section{Fabrication of the phase shifters}\label{appD}
\noindent For the phase plates we have used borosilicate glass of 180 $\mu$m and 1 mm thickness. All substrates have been polished and chemically cleaned prior to the coating runs. Anti-refractive coatings were deposited using a B\"uhler SyrusPro 1100 via ion-assisted deposition. The surface of the substrates was pre-treated with a gentle Ar-ion plasma at a flow of 11 sccm using 80 V bias-voltage for 100 s. This enhances the surface energy of the interface by cracking bonds of adsorbed hydrogen, thus creating chemically active sites, to which the subsequently deposited material may crosslink \cite{10.1364/OME.9.000598}. Additionally, the pre-treatment provides a further cleaning effect of the surface. In the next step, we coated the activated surface with a quarter-wave stack of \ce{MgF2} from an electron beam gun at a low deposition rate of 0.35 nm/s. The chamber temperature was increased to 300 $^\circ$C using four ceramic heaters to ensure a monocrystalline \ce{MgF2} layer. After the first coating run, the substrates were flipped and the procedure is repeated to also coat the backside. \ce{MgF2} was chosen due to its refractive index being in between that of air and borosilicate glass. The quarter-wave thickness for our single-photon wavelength is 307.9 nm. A characterization using a spectrophotometer before and after the coating showed that the reflection losses decreased from 8.39\% to 2.69\% at 565 nm.

\section{Matrix treatment of the interferometers}\label{appE}
\noindent Ideally, any interferometer has a visibility of unity. Realistically, however, the splitting ratios might not be mutually equivalent and each path has some loss, which reduces the interference contrast intrinsically. To characterize the coherence of a light source using an interferometer, this intrinsic limit serves as an upper bound the measurement cannot exceed and has to be calculated.\\
\indent An interferometer can be easily calculated using the transfer matrix formalism. The standard Mach-Zehnder interferometer (MZI) contains a beamsplitter with matrix
\begin{align}
\const{BS}=
\begin{pmatrix}
\sqrt{1-R}&\const{i}\sqrt{R}\\
\const{i}\sqrt{R} & \sqrt{1-R}\\
\end{pmatrix}
\end{align}
where $R$ is the reflection coefficient ($T=1-R$) and a matrix
\begin{align}
\const{P}=
\begin{pmatrix}
L_1\const{e}^{\const{i}\phi}&0\\
0 & L_2\\
\end{pmatrix}
\end{align}
describing the paths, where $\phi$ is a phase shift and $L_{1,2}$ contains all losses of each path, respectively. The total MZI can be described by concatenating the components:
\begin{align}
\const{MZI}=\const{BS}\times\const{P}\times\const{BS}
\end{align}
The output arms can be calculated by projecting the input multiplied by the MZI matrix. For the input
\begin{align}
\ket{\text{in}}=
\begin{pmatrix}
1\\
0\\
\end{pmatrix}
\end{align}
this gives the well-known $\sin^2\left(\phi\right)$ interference pattern. This relation also allows us to model our realistic beamsplitter with splitting ratios and losses described in Tab.\ \ref{tab:grating}, leading to the theoretical interference visibilities shown in Tab.\ \ref{tab:visibility}. Note that the MZI is equivalent to the Michelson interferometer in our experiment with the only difference that the phase shifter is passed twice, leading to an effective phase of $2\phi$.\\
\indent For the three-path interferometer the same approach can be used, extended to the third path. The beamsplitter matrix changes to that of a tritter, which can be constructed in general by concatenating three beamsplitters \cite{PhysRevA.55.2564}:
\begin{align}
\begin{gathered}
\begin{pmatrix}
\sqrt{1-R_1}&\const{i}\sqrt{R_1}&0\\
\const{i}\sqrt{R_1} & \sqrt{1-R_1}&0\\
0&0&1\\
\end{pmatrix}\times
\begin{pmatrix}
\sqrt{1-R_2}&0&\const{i}\sqrt{R_2}\\
0&1&0\\
\const{i}\sqrt{R_2} &0& \sqrt{1-R_2}\\
\end{pmatrix}\times\nonumber\\
\begin{pmatrix}
1&0&0\\
0&\sqrt{1-R_3}&\const{i}\sqrt{R_3}\\
0&\const{i}\sqrt{R_3} & \sqrt{1-R_3}\\
\end{pmatrix}
\end{gathered}
\end{align}
where $R_i$ are internal reflection coefficients. We calculate these coefficients numerically such that they represent our experimentally characterized tritter. The calculation of the three-path interferometer is then analogous to that of the MZI.

%\bibliography{bibliography}

\begin{thebibliography}{53}%
\makeatletter
\providecommand \@ifxundefined [1]{%
 \@ifx{#1\undefined}
}%
\providecommand \@ifnum [1]{%
 \ifnum #1\expandafter \@firstoftwo
 \else \expandafter \@secondoftwo
 \fi
}%
\providecommand \@ifx [1]{%
 \ifx #1\expandafter \@firstoftwo
 \else \expandafter \@secondoftwo
 \fi
}%
\providecommand \natexlab [1]{#1}%
\providecommand \enquote  [1]{``#1''}%
\providecommand \bibnamefont  [1]{#1}%
\providecommand \bibfnamefont [1]{#1}%
\providecommand \citenamefont [1]{#1}%
\providecommand \href@noop [0]{\@secondoftwo}%
\providecommand \href [0]{\begingroup \@sanitize@url \@href}%
\providecommand \@href[1]{\@@startlink{#1}\@@href}%
\providecommand \@@href[1]{\endgroup#1\@@endlink}%
\providecommand \@sanitize@url [0]{\catcode `\\12\catcode `\$12\catcode
  `\&12\catcode `\#12\catcode `\^12\catcode `\_12\catcode `\%12\relax}%
\providecommand \@@startlink[1]{}%
\providecommand \@@endlink[0]{}%
\providecommand \url  [0]{\begingroup\@sanitize@url \@url }%
\providecommand \@url [1]{\endgroup\@href {#1}{\urlprefix }}%
\providecommand \urlprefix  [0]{URL }%
\providecommand \Eprint [0]{\href }%
\providecommand \doibase [0]{http://dx.doi.org/}%
\providecommand \selectlanguage [0]{\@gobble}%
\providecommand \bibinfo  [0]{\@secondoftwo}%
\providecommand \bibfield  [0]{\@secondoftwo}%
\providecommand \translation [1]{[#1]}%
\providecommand \BibitemOpen [0]{}%
\providecommand \bibitemStop [0]{}%
\providecommand \bibitemNoStop [0]{.\EOS\space}%
\providecommand \EOS [0]{\spacefactor3000\relax}%
\providecommand \BibitemShut  [1]{\csname bibitem#1\endcsname}%
\let\auto@bib@innerbib\@empty
%</preamble>
\bibitem [{\citenamefont {Wang}\ \emph {et~al.}(2008)\citenamefont {Wang},
  \citenamefont {Song}, \citenamefont {Liu}, \citenamefont {Wu},\ and\
  \citenamefont {Fan}}]{10.1088/0957-4484/19/7/075609}%
  \BibitemOpen
  \bibfield  {author} {\bibinfo {author} {\bibfnamefont {Ding}\ \bibnamefont
  {Wang}}, \bibinfo {author} {\bibfnamefont {Pengcheng}\ \bibnamefont {Song}},
  \bibinfo {author} {\bibfnamefont {Changhong}\ \bibnamefont {Liu}}, \bibinfo
  {author} {\bibfnamefont {Wei}\ \bibnamefont {Wu}}, \ and\ \bibinfo {author}
  {\bibfnamefont {Shoushan}\ \bibnamefont {Fan}},\ }\bibfield  {title}
  {\enquote {\bibinfo {title} {Highly oriented carbon nanotube papers made of
  aligned carbon nanotubes},}\ }\href {\doibase 10.1088/0957-4484/19/7/075609}
  {\bibfield  {journal} {\bibinfo  {journal} {Nanotechnol.}\ }\textbf {\bibinfo
  {volume} {19}},\ \bibinfo {pages} {075609} (\bibinfo {year}
  {2008})}\BibitemShut {NoStop}%
\bibitem [{\citenamefont {Talapin}\ and\ \citenamefont
  {Shevchenko}(2016)}]{10.1021/acs.chemrev.6b00566}%
  \BibitemOpen
  \bibfield  {author} {\bibinfo {author} {\bibfnamefont {Dmitri~V.}\
  \bibnamefont {Talapin}}\ and\ \bibinfo {author} {\bibfnamefont {Elena~V.}\
  \bibnamefont {Shevchenko}},\ }\bibfield  {title} {\enquote {\bibinfo {title}
  {{Introduction: Nanoparticle Chemistry}},}\ }\href {\doibase
  10.1021/acs.chemrev.6b00566} {\bibfield  {journal} {\bibinfo  {journal}
  {Chem. Rev.}\ }\textbf {\bibinfo {volume} {116}},\ \bibinfo {pages}
  {10343--10345} (\bibinfo {year} {2016})}\BibitemShut {NoStop}%
\bibitem [{\citenamefont {Garnett}\ \emph {et~al.}(2019)\citenamefont
  {Garnett}, \citenamefont {Mai},\ and\ \citenamefont
  {Yang}}]{10.1021/acs.chemrev.9b00423}%
  \BibitemOpen
  \bibfield  {author} {\bibinfo {author} {\bibfnamefont {Erik}\ \bibnamefont
  {Garnett}}, \bibinfo {author} {\bibfnamefont {Liqiang}\ \bibnamefont {Mai}},
  \ and\ \bibinfo {author} {\bibfnamefont {Peidong}\ \bibnamefont {Yang}},\
  }\bibfield  {title} {\enquote {\bibinfo {title} {{Introduction: 1D
  Nanomaterials/Nanowires}},}\ }\href {\doibase 10.1021/acs.chemrev.9b00423}
  {\bibfield  {journal} {\bibinfo  {journal} {Chem. Rev.}\ }\textbf {\bibinfo
  {volume} {119}},\ \bibinfo {pages} {8955--8957} (\bibinfo {year}
  {2019})}\BibitemShut {NoStop}%
\bibitem [{\citenamefont {Allen}\ \emph {et~al.}(2010)\citenamefont {Allen},
  \citenamefont {Tung},\ and\ \citenamefont {Kaner}}]{10.1021/cr900070d}%
  \BibitemOpen
  \bibfield  {author} {\bibinfo {author} {\bibfnamefont {Matthew~J.}\
  \bibnamefont {Allen}}, \bibinfo {author} {\bibfnamefont {Vincent~C.}\
  \bibnamefont {Tung}}, \ and\ \bibinfo {author} {\bibfnamefont {Richard~B.}\
  \bibnamefont {Kaner}},\ }\bibfield  {title} {\enquote {\bibinfo {title}
  {{Honeycomb Carbon: A Review of Graphene}},}\ }\href {\doibase
  10.1021/cr900070d} {\bibfield  {journal} {\bibinfo  {journal} {Chem. Rev.}\
  }\textbf {\bibinfo {volume} {110}},\ \bibinfo {pages} {132--145} (\bibinfo
  {year} {2010})}\BibitemShut {NoStop}%
\bibitem [{\citenamefont {Manzeli}\ \emph {et~al.}(2017)\citenamefont
  {Manzeli}, \citenamefont {Ovchinnikov}, \citenamefont {Pasquier},
  \citenamefont {Yazyev},\ and\ \citenamefont
  {Kis}}]{10.1038/natrevmats.2017.33}%
  \BibitemOpen
  \bibfield  {author} {\bibinfo {author} {\bibfnamefont {Sajedeh}\ \bibnamefont
  {Manzeli}}, \bibinfo {author} {\bibfnamefont {Dmitry}\ \bibnamefont
  {Ovchinnikov}}, \bibinfo {author} {\bibfnamefont {Diego}\ \bibnamefont
  {Pasquier}}, \bibinfo {author} {\bibfnamefont {Oleg~V.}\ \bibnamefont
  {Yazyev}}, \ and\ \bibinfo {author} {\bibfnamefont {Andras}\ \bibnamefont
  {Kis}},\ }\bibfield  {title} {\enquote {\bibinfo {title} {{2D transition
  metal dichalcogenides}},}\ }\href {\doibase 10.1038/natrevmats.2017.33}
  {\bibfield  {journal} {\bibinfo  {journal} {Nat. Rev. Mater.}\ }\textbf
  {\bibinfo {volume} {2}},\ \bibinfo {pages} {17033} (\bibinfo {year}
  {2017})}\BibitemShut {NoStop}%
\bibitem [{\citenamefont {Xu}\ \emph {et~al.}(2018)\citenamefont {Xu},
  \citenamefont {Ma}, \citenamefont {Shen}, \citenamefont {Fatemi},
  \citenamefont {Wu}, \citenamefont {Chang}, \citenamefont {Chang},
  \citenamefont {Valdivia}, \citenamefont {Chan}, \citenamefont {Gibson},
  \citenamefont {Zhou}, \citenamefont {Liu}, \citenamefont {Watanabe},
  \citenamefont {Taniguchi}, \citenamefont {Lin}, \citenamefont {Cava},
  \citenamefont {Fu}, \citenamefont {Gedik},\ and\ \citenamefont
  {Jarillo-Herrero}}]{10.1038/s41567-018-0189-6}%
  \BibitemOpen
  \bibfield  {author} {\bibinfo {author} {\bibfnamefont {Su-Yang}\ \bibnamefont
  {Xu}}, \bibinfo {author} {\bibfnamefont {Qiong}\ \bibnamefont {Ma}}, \bibinfo
  {author} {\bibfnamefont {Huitao}\ \bibnamefont {Shen}}, \bibinfo {author}
  {\bibfnamefont {Valla}\ \bibnamefont {Fatemi}}, \bibinfo {author}
  {\bibfnamefont {Sanfeng}\ \bibnamefont {Wu}}, \bibinfo {author}
  {\bibfnamefont {Tay-Rong}\ \bibnamefont {Chang}}, \bibinfo {author}
  {\bibfnamefont {Guoqing}\ \bibnamefont {Chang}}, \bibinfo {author}
  {\bibfnamefont {Andr{\'e}s M.~Mier}\ \bibnamefont {Valdivia}}, \bibinfo
  {author} {\bibfnamefont {Ching-Kit}\ \bibnamefont {Chan}}, \bibinfo {author}
  {\bibfnamefont {Quinn~D.}\ \bibnamefont {Gibson}}, \bibinfo {author}
  {\bibfnamefont {Jiadong}\ \bibnamefont {Zhou}}, \bibinfo {author}
  {\bibfnamefont {Zheng}\ \bibnamefont {Liu}}, \bibinfo {author} {\bibfnamefont
  {Kenji}\ \bibnamefont {Watanabe}}, \bibinfo {author} {\bibfnamefont
  {Takashi}\ \bibnamefont {Taniguchi}}, \bibinfo {author} {\bibfnamefont
  {Hsin}\ \bibnamefont {Lin}}, \bibinfo {author} {\bibfnamefont {Robert~J.}\
  \bibnamefont {Cava}}, \bibinfo {author} {\bibfnamefont {Liang}\ \bibnamefont
  {Fu}}, \bibinfo {author} {\bibfnamefont {Nuh}\ \bibnamefont {Gedik}}, \ and\
  \bibinfo {author} {\bibfnamefont {Pablo}\ \bibnamefont {Jarillo-Herrero}},\
  }\bibfield  {title} {\enquote {\bibinfo {title} {{Electrically switchable
  Berry curvature dipole in the monolayer topological insulator WTe$_2$}},}\
  }\href {\doibase 10.1038/s41567-018-0189-6} {\bibfield  {journal} {\bibinfo
  {journal} {Nat. Phys.}\ }\textbf {\bibinfo {volume} {14}},\ \bibinfo {pages}
  {900--906} (\bibinfo {year} {2018})}\BibitemShut {NoStop}%
\bibitem [{\citenamefont {Wang}\ \emph {et~al.}(2015)\citenamefont {Wang},
  \citenamefont {Deng}, \citenamefont {Liu},\ and\ \citenamefont
  {Liu}}]{10.1093/nsr/nwu080}%
  \BibitemOpen
  \bibfield  {author} {\bibinfo {author} {\bibfnamefont {Jinying}\ \bibnamefont
  {Wang}}, \bibinfo {author} {\bibfnamefont {Shibin}\ \bibnamefont {Deng}},
  \bibinfo {author} {\bibfnamefont {Zhongfan}\ \bibnamefont {Liu}}, \ and\
  \bibinfo {author} {\bibfnamefont {Zhirong}\ \bibnamefont {Liu}},\ }\bibfield
  {title} {\enquote {\bibinfo {title} {{The rare two-dimensional materials with
  Dirac cones}},}\ }\href {\doibase 10.1093/nsr/nwu080} {\bibfield  {journal}
  {\bibinfo  {journal} {Natl. Sci. Rev.}\ }\textbf {\bibinfo {volume} {2}},\
  \bibinfo {pages} {22--39} (\bibinfo {year} {2015})}\BibitemShut {NoStop}%
\bibitem [{\citenamefont {Sharma}\ \emph {et~al.}(2020)\citenamefont {Sharma},
  \citenamefont {Zhang}, \citenamefont {Tollerud}, \citenamefont {Dong},
  \citenamefont {Zhu}, \citenamefont {Halbich}, \citenamefont {Vogl},
  \citenamefont {Liang}, \citenamefont {Nguyen}, \citenamefont {Wang},
  \citenamefont {Sanwlani}, \citenamefont {Earl}, \citenamefont {Macdonald},
  \citenamefont {Lam}, \citenamefont {Davis},\ and\ \citenamefont
  {Lu}}]{Sharma_2020}%
  \BibitemOpen
  \bibfield  {author} {\bibinfo {author} {\bibfnamefont {Ankur}\ \bibnamefont
  {Sharma}}, \bibinfo {author} {\bibfnamefont {Linglong}\ \bibnamefont
  {Zhang}}, \bibinfo {author} {\bibfnamefont {Jonathan~O.}\ \bibnamefont
  {Tollerud}}, \bibinfo {author} {\bibfnamefont {Miheng}\ \bibnamefont {Dong}},
  \bibinfo {author} {\bibfnamefont {Yi}~\bibnamefont {Zhu}}, \bibinfo {author}
  {\bibfnamefont {Robert}\ \bibnamefont {Halbich}}, \bibinfo {author}
  {\bibfnamefont {Tobias}\ \bibnamefont {Vogl}}, \bibinfo {author}
  {\bibfnamefont {Kun}\ \bibnamefont {Liang}}, \bibinfo {author} {\bibfnamefont
  {Hieu~T.}\ \bibnamefont {Nguyen}}, \bibinfo {author} {\bibfnamefont {Fan}\
  \bibnamefont {Wang}}, \bibinfo {author} {\bibfnamefont {Shilpa}\ \bibnamefont
  {Sanwlani}}, \bibinfo {author} {\bibfnamefont {Stuart~K.}\ \bibnamefont
  {Earl}}, \bibinfo {author} {\bibfnamefont {Daniel}\ \bibnamefont
  {Macdonald}}, \bibinfo {author} {\bibfnamefont {Ping~Koy}\ \bibnamefont
  {Lam}}, \bibinfo {author} {\bibfnamefont {Jeffrey~A.}\ \bibnamefont {Davis}},
  \ and\ \bibinfo {author} {\bibfnamefont {Yuerui}\ \bibnamefont {Lu}},\
  }\bibfield  {title} {\enquote {\bibinfo {title} {Supertransport of excitons
  in atomically thin organic semiconductors at the {2D} quantum limit},}\
  }\href {\doibase 10.1038/s41377-020-00347-y} {\bibfield  {journal} {\bibinfo
  {journal} {Light Sci. Appl.}\ }\textbf {\bibinfo {volume} {9}},\ \bibinfo
  {pages} {116} (\bibinfo {year} {2020})}\BibitemShut {NoStop}%
\bibitem [{\citenamefont {Mak}\ \emph {et~al.}(2010)\citenamefont {Mak},
  \citenamefont {Lee}, \citenamefont {Hone}, \citenamefont {Shan},\ and\
  \citenamefont {Heinz}}]{PhysRevLett.105.136805}%
  \BibitemOpen
  \bibfield  {author} {\bibinfo {author} {\bibfnamefont {Kin~Fai}\ \bibnamefont
  {Mak}}, \bibinfo {author} {\bibfnamefont {Changgu}\ \bibnamefont {Lee}},
  \bibinfo {author} {\bibfnamefont {James}\ \bibnamefont {Hone}}, \bibinfo
  {author} {\bibfnamefont {Jie}\ \bibnamefont {Shan}}, \ and\ \bibinfo {author}
  {\bibfnamefont {Tony~F.}\ \bibnamefont {Heinz}},\ }\bibfield  {title}
  {\enquote {\bibinfo {title} {{Atomically Thin ${\mathrm{MoS}}_{2}$: A New
  Direct-Gap Semiconductor}},}\ }\href {\doibase
  10.1103/PhysRevLett.105.136805} {\bibfield  {journal} {\bibinfo  {journal}
  {Phys. Rev. Lett.}\ }\textbf {\bibinfo {volume} {105}},\ \bibinfo {pages}
  {136805} (\bibinfo {year} {2010})}\BibitemShut {NoStop}%
\bibitem [{\citenamefont {Cao}\ \emph {et~al.}(2018)\citenamefont {Cao},
  \citenamefont {Fatemi}, \citenamefont {Fang}, \citenamefont {Watanabe},
  \citenamefont {Taniguchi}, \citenamefont {Kaxiras},\ and\ \citenamefont
  {Jarillo-Herrero}}]{10.1038/nature26160}%
  \BibitemOpen
  \bibfield  {author} {\bibinfo {author} {\bibfnamefont {Yuan}\ \bibnamefont
  {Cao}}, \bibinfo {author} {\bibfnamefont {Valla}\ \bibnamefont {Fatemi}},
  \bibinfo {author} {\bibfnamefont {Shiang}\ \bibnamefont {Fang}}, \bibinfo
  {author} {\bibfnamefont {Kenji}\ \bibnamefont {Watanabe}}, \bibinfo {author}
  {\bibfnamefont {Takashi}\ \bibnamefont {Taniguchi}}, \bibinfo {author}
  {\bibfnamefont {Efthimios}\ \bibnamefont {Kaxiras}}, \ and\ \bibinfo {author}
  {\bibfnamefont {Pablo}\ \bibnamefont {Jarillo-Herrero}},\ }\bibfield  {title}
  {\enquote {\bibinfo {title} {Unconventional superconductivity in magic-angle
  graphene superlattices},}\ }\href {\doibase 10.1038/nature26160} {\bibfield
  {journal} {\bibinfo  {journal} {Nature}\ }\textbf {\bibinfo {volume} {556}},\
  \bibinfo {pages} {43--50} (\bibinfo {year} {2018})}\BibitemShut {NoStop}%
\bibitem [{\citenamefont {Radisavljevic}\ \emph {et~al.}(2011)\citenamefont
  {Radisavljevic}, \citenamefont {Radenovic}, \citenamefont {Brivio},
  \citenamefont {Giacometti},\ and\ \citenamefont
  {Kis}}]{10.1038/nnano.2010.279}%
  \BibitemOpen
  \bibfield  {author} {\bibinfo {author} {\bibfnamefont {B.}~\bibnamefont
  {Radisavljevic}}, \bibinfo {author} {\bibfnamefont {A.}~\bibnamefont
  {Radenovic}}, \bibinfo {author} {\bibfnamefont {J.}~\bibnamefont {Brivio}},
  \bibinfo {author} {\bibfnamefont {V.}~\bibnamefont {Giacometti}}, \ and\
  \bibinfo {author} {\bibfnamefont {A.}~\bibnamefont {Kis}},\ }\bibfield
  {title} {\enquote {\bibinfo {title} {{Single-layer MoS$_2$ transistors}},}\
  }\href {\doibase 10.1038/nnano.2010.279} {\bibfield  {journal} {\bibinfo
  {journal} {Nat. Nanotechnol.}\ }\textbf {\bibinfo {volume} {6}},\ \bibinfo
  {pages} {147--150} (\bibinfo {year} {2011})}\BibitemShut {NoStop}%
\bibitem [{\citenamefont {Smolyanitsky}\ \emph {et~al.}(2016)\citenamefont
  {Smolyanitsky}, \citenamefont {Yakobson}, \citenamefont {Wassenaar},
  \citenamefont {Paulechka},\ and\ \citenamefont
  {Kroenlein}}]{10.1021/acsnano.6b05274}%
  \BibitemOpen
  \bibfield  {author} {\bibinfo {author} {\bibfnamefont {Alex}\ \bibnamefont
  {Smolyanitsky}}, \bibinfo {author} {\bibfnamefont {Boris~I.}\ \bibnamefont
  {Yakobson}}, \bibinfo {author} {\bibfnamefont {Tsjerk~A.}\ \bibnamefont
  {Wassenaar}}, \bibinfo {author} {\bibfnamefont {Eugene}\ \bibnamefont
  {Paulechka}}, \ and\ \bibinfo {author} {\bibfnamefont {Kenneth}\ \bibnamefont
  {Kroenlein}},\ }\bibfield  {title} {\enquote {\bibinfo {title} {{A
  MoS$_2$-Based Capacitive Displacement Sensor for DNA Sequencing}},}\ }\href
  {\doibase 10.1021/acsnano.6b05274} {\bibfield  {journal} {\bibinfo  {journal}
  {ACS Nano}\ }\textbf {\bibinfo {volume} {10}},\ \bibinfo {pages} {9009--9016}
  (\bibinfo {year} {2016})}\BibitemShut {NoStop}%
\bibitem [{\citenamefont {He}\ \emph {et~al.}(2012)\citenamefont {He},
  \citenamefont {Zeng}, \citenamefont {Yin}, \citenamefont {Li}, \citenamefont
  {Wu}, \citenamefont {Huang},\ and\ \citenamefont
  {Zhang}}]{10.1002/smll.201201224}%
  \BibitemOpen
  \bibfield  {author} {\bibinfo {author} {\bibfnamefont {Qiyuan}\ \bibnamefont
  {He}}, \bibinfo {author} {\bibfnamefont {Zhiyuan}\ \bibnamefont {Zeng}},
  \bibinfo {author} {\bibfnamefont {Zongyou}\ \bibnamefont {Yin}}, \bibinfo
  {author} {\bibfnamefont {Hai}\ \bibnamefont {Li}}, \bibinfo {author}
  {\bibfnamefont {Shixin}\ \bibnamefont {Wu}}, \bibinfo {author} {\bibfnamefont
  {Xiao}\ \bibnamefont {Huang}}, \ and\ \bibinfo {author} {\bibfnamefont {Hua}\
  \bibnamefont {Zhang}},\ }\bibfield  {title} {\enquote {\bibinfo {title}
  {{Fabrication of Flexible MoS2 Thin-Film Transistor Arrays for Practical
  Gas-Sensing Applications}},}\ }\href {\doibase 10.1002/smll.201201224}
  {\bibfield  {journal} {\bibinfo  {journal} {Small}\ }\textbf {\bibinfo
  {volume} {8}},\ \bibinfo {pages} {2994--2999} (\bibinfo {year}
  {2012})}\BibitemShut {NoStop}%
\bibitem [{\citenamefont {Vogl}\ \emph
  {et~al.}(2019{\natexlab{a}})\citenamefont {Vogl}, \citenamefont {Sripathy},
  \citenamefont {Sharma}, \citenamefont {Reddy}, \citenamefont {Sullivan},
  \citenamefont {Machacek}, \citenamefont {Zhang}, \citenamefont {Karouta},
  \citenamefont {Buchler}, \citenamefont {Doherty}, \citenamefont {Lu},\ and\
  \citenamefont {Lam}}]{10.1038/s41467-019-09219-5}%
  \BibitemOpen
  \bibfield  {author} {\bibinfo {author} {\bibfnamefont {T.}~\bibnamefont
  {Vogl}}, \bibinfo {author} {\bibfnamefont {K.}~\bibnamefont {Sripathy}},
  \bibinfo {author} {\bibfnamefont {A.}~\bibnamefont {Sharma}}, \bibinfo
  {author} {\bibfnamefont {P.}~\bibnamefont {Reddy}}, \bibinfo {author}
  {\bibfnamefont {J.}~\bibnamefont {Sullivan}}, \bibinfo {author}
  {\bibfnamefont {J.~R.}\ \bibnamefont {Machacek}}, \bibinfo {author}
  {\bibfnamefont {L.}~\bibnamefont {Zhang}}, \bibinfo {author} {\bibfnamefont
  {F.}~\bibnamefont {Karouta}}, \bibinfo {author} {\bibfnamefont {B.~C.}\
  \bibnamefont {Buchler}}, \bibinfo {author} {\bibfnamefont {M.~W.}\
  \bibnamefont {Doherty}}, \bibinfo {author} {\bibfnamefont {Y.}~\bibnamefont
  {Lu}}, \ and\ \bibinfo {author} {\bibfnamefont {P.~K.}\ \bibnamefont {Lam}},\
  }\bibfield  {title} {\enquote {\bibinfo {title} {Radiation tolerance of
  two-dimensional material-based devices for space applications},}\ }\href
  {\doibase 10.1038/s41467-019-09219-5} {\bibfield  {journal} {\bibinfo
  {journal} {Nat. Commun.}\ }\textbf {\bibinfo {volume} {10}},\ \bibinfo
  {pages} {1202} (\bibinfo {year} {2019}{\natexlab{a}})}\BibitemShut {NoStop}%
\bibitem [{\citenamefont {Liu}\ \emph {et~al.}(2017)\citenamefont {Liu},
  \citenamefont {Shivananju}, \citenamefont {Wang}, \citenamefont {Zhang},
  \citenamefont {Yu}, \citenamefont {Xiao}, \citenamefont {Sun}, \citenamefont
  {Ma}, \citenamefont {Mu}, \citenamefont {Lin}, \citenamefont {Zhang},
  \citenamefont {Lu}, \citenamefont {Qiu}, \citenamefont {Li},\ and\
  \citenamefont {Bao}}]{10.1021/acsami.7b09889}%
  \BibitemOpen
  \bibfield  {author} {\bibinfo {author} {\bibfnamefont {Yan}\ \bibnamefont
  {Liu}}, \bibinfo {author} {\bibfnamefont {Bannur~Nanjunda}\ \bibnamefont
  {Shivananju}}, \bibinfo {author} {\bibfnamefont {Yusheng}\ \bibnamefont
  {Wang}}, \bibinfo {author} {\bibfnamefont {Yupeng}\ \bibnamefont {Zhang}},
  \bibinfo {author} {\bibfnamefont {Wenzhi}\ \bibnamefont {Yu}}, \bibinfo
  {author} {\bibfnamefont {Si}~\bibnamefont {Xiao}}, \bibinfo {author}
  {\bibfnamefont {Tian}\ \bibnamefont {Sun}}, \bibinfo {author} {\bibfnamefont
  {Weiliang}\ \bibnamefont {Ma}}, \bibinfo {author} {\bibfnamefont {Haoran}\
  \bibnamefont {Mu}}, \bibinfo {author} {\bibfnamefont {Shenghuang}\
  \bibnamefont {Lin}}, \bibinfo {author} {\bibfnamefont {Han}\ \bibnamefont
  {Zhang}}, \bibinfo {author} {\bibfnamefont {Yuerui}\ \bibnamefont {Lu}},
  \bibinfo {author} {\bibfnamefont {Cheng-Wei}\ \bibnamefont {Qiu}}, \bibinfo
  {author} {\bibfnamefont {Shaojuan}\ \bibnamefont {Li}}, \ and\ \bibinfo
  {author} {\bibfnamefont {Qiaoliang}\ \bibnamefont {Bao}},\ }\bibfield
  {title} {\enquote {\bibinfo {title} {{Highly Efficient and Air-Stable
  Infrared Photodetector Based on 2D Layered Graphene–Black Phosphorus
  Heterostructure}},}\ }\href {\doibase 10.1021/acsami.7b09889} {\bibfield
  {journal} {\bibinfo  {journal} {ACS Appl. Mater. Interfaces}\ }\textbf
  {\bibinfo {volume} {9}},\ \bibinfo {pages} {36137--36145} (\bibinfo {year}
  {2017})}\BibitemShut {NoStop}%
\bibitem [{\citenamefont {Born}(1926)}]{10.1007/BF01397184}%
  \BibitemOpen
  \bibfield  {author} {\bibinfo {author} {\bibfnamefont {Max}\ \bibnamefont
  {Born}},\ }\bibfield  {title} {\enquote {\bibinfo {title} {{Zur
  Quantenmechanik der Sto\ss vorg\"ange.}}}\ }\href {\doibase
  10.1007/BF01397184} {\bibfield  {journal} {\bibinfo  {journal} {Z. Physik}\
  }\textbf {\bibinfo {volume} {38}},\ \bibinfo {pages} {803--827} (\bibinfo
  {year} {1926})}\BibitemShut {NoStop}%
\bibitem [{\citenamefont {Mould}(2005)}]{arXiv:quant-ph/0507170}%
  \BibitemOpen
  \bibfield  {author} {\bibinfo {author} {\bibfnamefont {Richard~A.}\
  \bibnamefont {Mould}},\ }\bibfield  {title} {\enquote {\bibinfo {title}
  {Without the born rule},}\ }\href {https://arxiv.org/abs/quant-ph/0507170}
  {\bibfield  {journal} {\bibinfo  {journal} {arXiv:quant-ph/0507170}\ }
  (\bibinfo {year} {2005})}\BibitemShut {NoStop}%
\bibitem [{\citenamefont {Saunders}(2004)}]{doi:10.1098/rspa.2003.1230}%
  \BibitemOpen
  \bibfield  {author} {\bibinfo {author} {\bibfnamefont {Simon}\ \bibnamefont
  {Saunders}},\ }\bibfield  {title} {\enquote {\bibinfo {title} {{Derivation of
  the Born rule from operational assumptions}},}\ }\href {\doibase
  10.1098/rspa.2003.1230} {\bibfield  {journal} {\bibinfo  {journal} {Proc.:
  Math., Phys. Eng. Sci.}\ }\textbf {\bibinfo {volume} {460}},\ \bibinfo
  {pages} {1771--1788} (\bibinfo {year} {2004})}\BibitemShut {NoStop}%
\bibitem [{\citenamefont {Gleason}(1957)}]{10.2307/24900629}%
  \BibitemOpen
  \bibfield  {author} {\bibinfo {author} {\bibfnamefont {Andrew~M.}\
  \bibnamefont {Gleason}},\ }\bibfield  {title} {\enquote {\bibinfo {title}
  {Measures on the closed subspaces of a hilbert space},}\ }\href
  {http://www.jstor.org/stable/24900629} {\bibfield  {journal} {\bibinfo
  {journal} {J. Math. Mech.}\ }\textbf {\bibinfo {volume} {6}},\ \bibinfo
  {pages} {885--893} (\bibinfo {year} {1957})}\BibitemShut {NoStop}%
\bibitem [{\citenamefont {Cooke}\ \emph {et~al.}(1985)\citenamefont {Cooke},
  \citenamefont {Keane},\ and\ \citenamefont {Moran}}]{cooke_keane_moran_1985}%
  \BibitemOpen
  \bibfield  {author} {\bibinfo {author} {\bibfnamefont {Roger}\ \bibnamefont
  {Cooke}}, \bibinfo {author} {\bibfnamefont {Michael}\ \bibnamefont {Keane}},
  \ and\ \bibinfo {author} {\bibfnamefont {William}\ \bibnamefont {Moran}},\
  }\bibfield  {title} {\enquote {\bibinfo {title} {An elementary proof of
  gleason's theorem},}\ }\href {\doibase 10.1017/S0305004100063313} {\bibfield
  {journal} {\bibinfo  {journal} {Math. Proc. Camb. Philos. Soc.}\ }\textbf
  {\bibinfo {volume} {98}},\ \bibinfo {pages} {117–128} (\bibinfo {year}
  {1985})}\BibitemShut {NoStop}%
\bibitem [{\citenamefont {Masanes}\ \emph {et~al.}(2019)\citenamefont
  {Masanes}, \citenamefont {Galley},\ and\ \citenamefont
  {M{\"u}ller}}]{Masanes2019}%
  \BibitemOpen
  \bibfield  {author} {\bibinfo {author} {\bibfnamefont {Llu{\'i}s}\
  \bibnamefont {Masanes}}, \bibinfo {author} {\bibfnamefont {Thomas~D.}\
  \bibnamefont {Galley}}, \ and\ \bibinfo {author} {\bibfnamefont {Markus~P.}\
  \bibnamefont {M{\"u}ller}},\ }\bibfield  {title} {\enquote {\bibinfo {title}
  {The measurement postulates of quantum mechanics are operationally
  redundant},}\ }\href {\doibase 10.1038/s41467-019-09348-x} {\bibfield
  {journal} {\bibinfo  {journal} {Nature Commun.}\ }\textbf {\bibinfo {volume}
  {10}},\ \bibinfo {pages} {1361} (\bibinfo {year} {2019})}\BibitemShut
  {NoStop}%
\bibitem [{\citenamefont {Sorkin}(1994)}]{10.1142/S021773239400294X}%
  \BibitemOpen
  \bibfield  {author} {\bibinfo {author} {\bibfnamefont {Rafael~D.}\
  \bibnamefont {Sorkin}},\ }\bibfield  {title} {\enquote {\bibinfo {title}
  {{Quantum Mechanics as Quantum Measure Theory}},}\ }\href {\doibase
  10.1142/S021773239400294X} {\bibfield  {journal} {\bibinfo  {journal} {Mod.
  Phys. Lett. A}\ }\textbf {\bibinfo {volume} {9}},\ \bibinfo {pages}
  {3119--3127} (\bibinfo {year} {1994})}\BibitemShut {NoStop}%
\bibitem [{\citenamefont {Peres}(1979)}]{PhysRevLett.42.683}%
  \BibitemOpen
  \bibfield  {author} {\bibinfo {author} {\bibfnamefont {Asher}\ \bibnamefont
  {Peres}},\ }\bibfield  {title} {\enquote {\bibinfo {title} {Proposed test for
  complex versus quaternion quantum theory},}\ }\href {\doibase
  10.1103/PhysRevLett.42.683} {\bibfield  {journal} {\bibinfo  {journal} {Phys.
  Rev. Lett.}\ }\textbf {\bibinfo {volume} {42}},\ \bibinfo {pages} {683--686}
  (\bibinfo {year} {1979})}\BibitemShut {NoStop}%
\bibitem [{\citenamefont {Kaiser}\ \emph {et~al.}(1984)\citenamefont {Kaiser},
  \citenamefont {George},\ and\ \citenamefont {Werner}}]{PhysRevA.29.2276}%
  \BibitemOpen
  \bibfield  {author} {\bibinfo {author} {\bibfnamefont {H.}~\bibnamefont
  {Kaiser}}, \bibinfo {author} {\bibfnamefont {E.~A.}\ \bibnamefont {George}},
  \ and\ \bibinfo {author} {\bibfnamefont {S.~A.}\ \bibnamefont {Werner}},\
  }\bibfield  {title} {\enquote {\bibinfo {title} {Neutron interferometric
  search for quaternions in quantum mechanics},}\ }\href {\doibase
  10.1103/PhysRevA.29.2276} {\bibfield  {journal} {\bibinfo  {journal} {Phys.
  Rev. A}\ }\textbf {\bibinfo {volume} {29}},\ \bibinfo {pages} {2276--2279}
  (\bibinfo {year} {1984})}\BibitemShut {NoStop}%
\bibitem [{\citenamefont {Procopio}\ \emph {et~al.}(2017)\citenamefont
  {Procopio}, \citenamefont {Rozema}, \citenamefont {Wong}, \citenamefont
  {Hamel}, \citenamefont {O'Brien}, \citenamefont {Zhang}, \citenamefont
  {Daki{\'{c}}},\ and\ \citenamefont {Walther}}]{10.1038/ncomms15044}%
  \BibitemOpen
  \bibfield  {author} {\bibinfo {author} {\bibfnamefont {Lorenzo~M.}\
  \bibnamefont {Procopio}}, \bibinfo {author} {\bibfnamefont {Lee~A.}\
  \bibnamefont {Rozema}}, \bibinfo {author} {\bibfnamefont {Zi~Jing}\
  \bibnamefont {Wong}}, \bibinfo {author} {\bibfnamefont {Deny~R.}\
  \bibnamefont {Hamel}}, \bibinfo {author} {\bibfnamefont {Kevin}\ \bibnamefont
  {O'Brien}}, \bibinfo {author} {\bibfnamefont {Xiang}\ \bibnamefont {Zhang}},
  \bibinfo {author} {\bibfnamefont {Borivoje}\ \bibnamefont {Daki{\'{c}}}}, \
  and\ \bibinfo {author} {\bibfnamefont {Philip}\ \bibnamefont {Walther}},\
  }\bibfield  {title} {\enquote {\bibinfo {title} {Single-photon test of
  hyper-complex quantum theories using a metamaterial},}\ }\href {\doibase
  10.1038/ncomms15044} {\bibfield  {journal} {\bibinfo  {journal} {Nat.
  Commun.}\ }\textbf {\bibinfo {volume} {8}},\ \bibinfo {pages} {15044}
  (\bibinfo {year} {2017})}\BibitemShut {NoStop}%
\bibitem [{\citenamefont {Sinha}\ \emph {et~al.}(2010)\citenamefont {Sinha},
  \citenamefont {Couteau}, \citenamefont {Jennewein}, \citenamefont
  {Laflamme},\ and\ \citenamefont {Weihs}}]{10.1126/science.1190545}%
  \BibitemOpen
  \bibfield  {author} {\bibinfo {author} {\bibfnamefont {Urbasi}\ \bibnamefont
  {Sinha}}, \bibinfo {author} {\bibfnamefont {Christophe}\ \bibnamefont
  {Couteau}}, \bibinfo {author} {\bibfnamefont {Thomas}\ \bibnamefont
  {Jennewein}}, \bibinfo {author} {\bibfnamefont {Raymond}\ \bibnamefont
  {Laflamme}}, \ and\ \bibinfo {author} {\bibfnamefont {Gregor}\ \bibnamefont
  {Weihs}},\ }\bibfield  {title} {\enquote {\bibinfo {title} {Ruling out
  multi-order interference in quantum mechanics},}\ }\href {\doibase
  10.1126/science.1190545} {\bibfield  {journal} {\bibinfo  {journal}
  {Science}\ }\textbf {\bibinfo {volume} {329}},\ \bibinfo {pages} {418--421}
  (\bibinfo {year} {2010})}\BibitemShut {NoStop}%
\bibitem [{\citenamefont {S{\"o}llner}\ \emph {et~al.}(2012)\citenamefont
  {S{\"o}llner}, \citenamefont {Gsch{\"o}sser}, \citenamefont {Mai},
  \citenamefont {Pressl}, \citenamefont {V{\"o}r{\"o}s},\ and\ \citenamefont
  {Weihs}}]{10.1007/s10701-011-9597-5}%
  \BibitemOpen
  \bibfield  {author} {\bibinfo {author} {\bibfnamefont {Immo}\ \bibnamefont
  {S{\"o}llner}}, \bibinfo {author} {\bibfnamefont {Benjamin}\ \bibnamefont
  {Gsch{\"o}sser}}, \bibinfo {author} {\bibfnamefont {Patrick}\ \bibnamefont
  {Mai}}, \bibinfo {author} {\bibfnamefont {Benedikt}\ \bibnamefont {Pressl}},
  \bibinfo {author} {\bibfnamefont {Zolt{\'a}n}\ \bibnamefont {V{\"o}r{\"o}s}},
  \ and\ \bibinfo {author} {\bibfnamefont {Gregor}\ \bibnamefont {Weihs}},\
  }\bibfield  {title} {\enquote {\bibinfo {title} {Testing born's rule in
  quantum mechanics for three mutually exclusive events},}\ }\href {\doibase
  10.1007/s10701-011-9597-5} {\bibfield  {journal} {\bibinfo  {journal} {Found.
  Phys.}\ }\textbf {\bibinfo {volume} {42}},\ \bibinfo {pages} {742--751}
  (\bibinfo {year} {2012})}\BibitemShut {NoStop}%
\bibitem [{\citenamefont {Kauten}\ \emph {et~al.}(2017)\citenamefont {Kauten},
  \citenamefont {Keil}, \citenamefont {Kaufmann}, \citenamefont {Pressl},
  \citenamefont {Brukner},\ and\ \citenamefont
  {Weihs}}]{10.1088/1367-2630/aa5d98}%
  \BibitemOpen
  \bibfield  {author} {\bibinfo {author} {\bibfnamefont {Thomas}\ \bibnamefont
  {Kauten}}, \bibinfo {author} {\bibfnamefont {Robert}\ \bibnamefont {Keil}},
  \bibinfo {author} {\bibfnamefont {Thomas}\ \bibnamefont {Kaufmann}}, \bibinfo
  {author} {\bibfnamefont {Benedikt}\ \bibnamefont {Pressl}}, \bibinfo {author}
  {\bibfnamefont {{\v{C}}aslav}\ \bibnamefont {Brukner}}, \ and\ \bibinfo
  {author} {\bibfnamefont {Gregor}\ \bibnamefont {Weihs}},\ }\bibfield  {title}
  {\enquote {\bibinfo {title} {Obtaining tight bounds on higher-order
  interferences with a 5-path interferometer},}\ }\href {\doibase
  10.1088/1367-2630/aa5d98} {\bibfield  {journal} {\bibinfo  {journal} {New J.
  Phys.}\ }\textbf {\bibinfo {volume} {19}},\ \bibinfo {pages} {033017}
  (\bibinfo {year} {2017})}\BibitemShut {NoStop}%
\bibitem [{\citenamefont {Pleinert}\ \emph
  {et~al.}(2020{\natexlab{a}})\citenamefont {Pleinert}, \citenamefont {von
  Zanthier},\ and\ \citenamefont {Lutz}}]{PhysRevResearch.2.012051}%
  \BibitemOpen
  \bibfield  {author} {\bibinfo {author} {\bibfnamefont {Marc-Oliver}\
  \bibnamefont {Pleinert}}, \bibinfo {author} {\bibfnamefont {Joachim}\
  \bibnamefont {von Zanthier}}, \ and\ \bibinfo {author} {\bibfnamefont {Eric}\
  \bibnamefont {Lutz}},\ }\bibfield  {title} {\enquote {\bibinfo {title}
  {Many-particle interference to test born's rule},}\ }\href {\doibase
  10.1103/PhysRevResearch.2.012051} {\bibfield  {journal} {\bibinfo  {journal}
  {Phys. Rev. Research}\ }\textbf {\bibinfo {volume} {2}},\ \bibinfo {pages}
  {012051} (\bibinfo {year} {2020}{\natexlab{a}})}\BibitemShut {NoStop}%
\bibitem [{\citenamefont {Pleinert}\ \emph
  {et~al.}(2020{\natexlab{b}})\citenamefont {Pleinert}, \citenamefont {Rueda},
  \citenamefont {Lutz},\ and\ \citenamefont {von Zanthier}}]{arXiv:2006.09496}%
  \BibitemOpen
  \bibfield  {author} {\bibinfo {author} {\bibfnamefont {Marc-Oliver}\
  \bibnamefont {Pleinert}}, \bibinfo {author} {\bibfnamefont {Alfredo}\
  \bibnamefont {Rueda}}, \bibinfo {author} {\bibfnamefont {Eric}\ \bibnamefont
  {Lutz}}, \ and\ \bibinfo {author} {\bibfnamefont {Joachim}\ \bibnamefont {von
  Zanthier}},\ }\bibfield  {title} {\enquote {\bibinfo {title} {Testing quantum
  theory with higher-order interference in many-particle correlations},}\
  }\href {https://arxiv.org/abs/2006.09496} {\bibfield  {journal} {\bibinfo
  {journal} {arXiv:2006.09496}\ } (\bibinfo {year}
  {2020}{\natexlab{b}})}\BibitemShut {NoStop}%
\bibitem [{\citenamefont {Vogl}\ \emph
  {et~al.}(2019{\natexlab{b}})\citenamefont {Vogl}, \citenamefont {Lecamwasam},
  \citenamefont {Buchler}, \citenamefont {Lu},\ and\ \citenamefont
  {Lam}}]{10.1021/acsphotonics.9b00314}%
  \BibitemOpen
  \bibfield  {author} {\bibinfo {author} {\bibfnamefont {Tobias}\ \bibnamefont
  {Vogl}}, \bibinfo {author} {\bibfnamefont {Ruvi}\ \bibnamefont {Lecamwasam}},
  \bibinfo {author} {\bibfnamefont {Ben~C.}\ \bibnamefont {Buchler}}, \bibinfo
  {author} {\bibfnamefont {Yuerui}\ \bibnamefont {Lu}}, \ and\ \bibinfo
  {author} {\bibfnamefont {Ping~Koy}\ \bibnamefont {Lam}},\ }\bibfield  {title}
  {\enquote {\bibinfo {title} {Compact cavity-enhanced single-photon generation
  with hexagonal boron nitride},}\ }\href {\doibase
  10.1021/acsphotonics.9b00314} {\bibfield  {journal} {\bibinfo  {journal} {ACS
  Photonics}\ }\textbf {\bibinfo {volume} {6}},\ \bibinfo {pages} {1955--1962}
  (\bibinfo {year} {2019}{\natexlab{b}})}\BibitemShut {NoStop}%
\bibitem [{\citenamefont {Kauten}\ \emph {et~al.}(2014)\citenamefont {Kauten},
  \citenamefont {Pressl}, \citenamefont {Kaufmann},\ and\ \citenamefont
  {Weihs}}]{10.1063/1.4879820}%
  \BibitemOpen
  \bibfield  {author} {\bibinfo {author} {\bibfnamefont {Thomas}\ \bibnamefont
  {Kauten}}, \bibinfo {author} {\bibfnamefont {Benedikt}\ \bibnamefont
  {Pressl}}, \bibinfo {author} {\bibfnamefont {Thomas}\ \bibnamefont
  {Kaufmann}}, \ and\ \bibinfo {author} {\bibfnamefont {Gregor}\ \bibnamefont
  {Weihs}},\ }\bibfield  {title} {\enquote {\bibinfo {title} {Measurement and
  modeling of the nonlinearity of photovoltaic and geiger-mode photodiodes},}\
  }\href {\doibase 10.1063/1.4879820} {\bibfield  {journal} {\bibinfo
  {journal} {Rev. Sci. Instrum.}\ }\textbf {\bibinfo {volume} {85}},\ \bibinfo
  {pages} {063102} (\bibinfo {year} {2014})}\BibitemShut {NoStop}%
\bibitem [{\citenamefont {Tran}\ \emph
  {et~al.}(2016{\natexlab{a}})\citenamefont {Tran}, \citenamefont {Bray},
  \citenamefont {Ford}, \citenamefont {Toth},\ and\ \citenamefont
  {Aharonovich}}]{nnano.2015.242}%
  \BibitemOpen
  \bibfield  {author} {\bibinfo {author} {\bibfnamefont {T.~T.}\ \bibnamefont
  {Tran}}, \bibinfo {author} {\bibfnamefont {K.}~\bibnamefont {Bray}}, \bibinfo
  {author} {\bibfnamefont {M.~J.}\ \bibnamefont {Ford}}, \bibinfo {author}
  {\bibfnamefont {M.}~\bibnamefont {Toth}}, \ and\ \bibinfo {author}
  {\bibfnamefont {I.}~\bibnamefont {Aharonovich}},\ }\bibfield  {title}
  {\enquote {\bibinfo {title} {Quantum emission from hexagonal boron nitride
  monolayers},}\ }\href {\doibase 10.1038/nnano.2015.242} {\bibfield  {journal}
  {\bibinfo  {journal} {Nat. Nanotechnol.}\ }\textbf {\bibinfo {volume} {11}},\
  \bibinfo {pages} {37--41} (\bibinfo {year} {2016}{\natexlab{a}})}\BibitemShut
  {NoStop}%
\bibitem [{\citenamefont {Tran}\ \emph
  {et~al.}(2016{\natexlab{b}})\citenamefont {Tran}, \citenamefont {Elbadawi},
  \citenamefont {Totonjian}, \citenamefont {Lobo}, \citenamefont {Grosso},
  \citenamefont {Moon}, \citenamefont {Englund}, \citenamefont {Ford},
  \citenamefont {Aharonovich},\ and\ \citenamefont
  {Toth}}]{10.1021/acsnano.6b03602}%
  \BibitemOpen
  \bibfield  {author} {\bibinfo {author} {\bibfnamefont {Toan~Trong}\
  \bibnamefont {Tran}}, \bibinfo {author} {\bibfnamefont {Christopher}\
  \bibnamefont {Elbadawi}}, \bibinfo {author} {\bibfnamefont {Daniel}\
  \bibnamefont {Totonjian}}, \bibinfo {author} {\bibfnamefont {Charlene~J.}\
  \bibnamefont {Lobo}}, \bibinfo {author} {\bibfnamefont {Gabriele}\
  \bibnamefont {Grosso}}, \bibinfo {author} {\bibfnamefont {Hyowon}\
  \bibnamefont {Moon}}, \bibinfo {author} {\bibfnamefont {Dirk~R.}\
  \bibnamefont {Englund}}, \bibinfo {author} {\bibfnamefont {Michael~J.}\
  \bibnamefont {Ford}}, \bibinfo {author} {\bibfnamefont {Igor}\ \bibnamefont
  {Aharonovich}}, \ and\ \bibinfo {author} {\bibfnamefont {Milos}\ \bibnamefont
  {Toth}},\ }\bibfield  {title} {\enquote {\bibinfo {title} {Robust multicolor
  single photon emission from point defects in hexagonal boron nitride},}\
  }\href {\doibase 10.1021/acsnano.6b03602} {\bibfield  {journal} {\bibinfo
  {journal} {ACS Nano}\ }\textbf {\bibinfo {volume} {10}},\ \bibinfo {pages}
  {7331--7338} (\bibinfo {year} {2016}{\natexlab{b}})}\BibitemShut {NoStop}%
\bibitem [{\citenamefont {Vogl}\ \emph {et~al.}(2017)\citenamefont {Vogl},
  \citenamefont {Lu},\ and\ \citenamefont {Lam}}]{10.1088/1361-6463/aa7839}%
  \BibitemOpen
  \bibfield  {author} {\bibinfo {author} {\bibfnamefont {Tobias}\ \bibnamefont
  {Vogl}}, \bibinfo {author} {\bibfnamefont {Yuerui}\ \bibnamefont {Lu}}, \
  and\ \bibinfo {author} {\bibfnamefont {Ping~Koy}\ \bibnamefont {Lam}},\
  }\bibfield  {title} {\enquote {\bibinfo {title} {Room temperature single
  photon source using fiber-integrated hexagonal boron nitride},}\ }\href
  {https://doi.org/10.1088/1361-6463/aa7839} {\bibfield  {journal} {\bibinfo
  {journal} {J. Phys. D: Appl. Phys.}\ }\textbf {\bibinfo {volume} {50}},\
  \bibinfo {pages} {295101} (\bibinfo {year} {2017})}\BibitemShut {NoStop}%
\bibitem [{\citenamefont {Vogl}\ \emph {et~al.}(2018)\citenamefont {Vogl},
  \citenamefont {Campbell}, \citenamefont {Buchler}, \citenamefont {Lu},\ and\
  \citenamefont {Lam}}]{10.1021/acsphotonics.8b00127}%
  \BibitemOpen
  \bibfield  {author} {\bibinfo {author} {\bibfnamefont {Tobias}\ \bibnamefont
  {Vogl}}, \bibinfo {author} {\bibfnamefont {Geoff}\ \bibnamefont {Campbell}},
  \bibinfo {author} {\bibfnamefont {Ben~C.}\ \bibnamefont {Buchler}}, \bibinfo
  {author} {\bibfnamefont {Yuerui}\ \bibnamefont {Lu}}, \ and\ \bibinfo
  {author} {\bibfnamefont {Ping~Koy}\ \bibnamefont {Lam}},\ }\bibfield  {title}
  {\enquote {\bibinfo {title} {Fabrication and deterministic transfer of
  high-quality quantum emitters in hexagonal boron nitride},}\ }\href {\doibase
  10.1021/acsphotonics.8b00127} {\bibfield  {journal} {\bibinfo  {journal} {ACS
  Photonics}\ }\textbf {\bibinfo {volume} {5}},\ \bibinfo {pages} {2305--2312}
  (\bibinfo {year} {2018})}\BibitemShut {NoStop}%
\bibitem [{\citenamefont {Nikolay}\ \emph {et~al.}(2019)\citenamefont
  {Nikolay}, \citenamefont {Mendelson}, \citenamefont {\"{O}zelci},
  \citenamefont {Sontheimer}, \citenamefont {B\"{o}hm}, \citenamefont {Kewes},
  \citenamefont {Toth}, \citenamefont {Aharonovich},\ and\ \citenamefont
  {Benson}}]{10.1364/OPTICA.6.001084}%
  \BibitemOpen
  \bibfield  {author} {\bibinfo {author} {\bibfnamefont {Niko}\ \bibnamefont
  {Nikolay}}, \bibinfo {author} {\bibfnamefont {Noah}\ \bibnamefont
  {Mendelson}}, \bibinfo {author} {\bibfnamefont {Ersan}\ \bibnamefont
  {\"{O}zelci}}, \bibinfo {author} {\bibfnamefont {Bernd}\ \bibnamefont
  {Sontheimer}}, \bibinfo {author} {\bibfnamefont {Florian}\ \bibnamefont
  {B\"{o}hm}}, \bibinfo {author} {\bibfnamefont {G\"{u}nter}\ \bibnamefont
  {Kewes}}, \bibinfo {author} {\bibfnamefont {Milos}\ \bibnamefont {Toth}},
  \bibinfo {author} {\bibfnamefont {Igor}\ \bibnamefont {Aharonovich}}, \ and\
  \bibinfo {author} {\bibfnamefont {Oliver}\ \bibnamefont {Benson}},\
  }\bibfield  {title} {\enquote {\bibinfo {title} {Direct measurement of
  quantum efficiency of single-photon emitters in hexagonal boron nitride},}\
  }\href {http://doi.org/10.1364/OPTICA.6.001084} {\bibfield  {journal}
  {\bibinfo  {journal} {Optica}\ }\textbf {\bibinfo {volume} {6}},\ \bibinfo
  {pages} {1084--1088} (\bibinfo {year} {2019})}\BibitemShut {NoStop}%
\bibitem [{\citenamefont {Vogl}\ \emph
  {et~al.}(2019{\natexlab{c}})\citenamefont {Vogl}, \citenamefont {Doherty},
  \citenamefont {Buchler}, \citenamefont {Lu},\ and\ \citenamefont
  {Lam}}]{C9NR04269E}%
  \BibitemOpen
  \bibfield  {author} {\bibinfo {author} {\bibfnamefont {Tobias}\ \bibnamefont
  {Vogl}}, \bibinfo {author} {\bibfnamefont {Marcus~W.}\ \bibnamefont
  {Doherty}}, \bibinfo {author} {\bibfnamefont {Ben~C.}\ \bibnamefont
  {Buchler}}, \bibinfo {author} {\bibfnamefont {Yuerui}\ \bibnamefont {Lu}}, \
  and\ \bibinfo {author} {\bibfnamefont {Ping~Koy}\ \bibnamefont {Lam}},\
  }\bibfield  {title} {\enquote {\bibinfo {title} {Atomic localization of
  quantum emitters in multilayer hexagonal boron nitride},}\ }\href {\doibase
  10.1039/C9NR04269E} {\bibfield  {journal} {\bibinfo  {journal} {Nanoscale}\
  }\textbf {\bibinfo {volume} {11}},\ \bibinfo {pages} {14362--14371} (\bibinfo
  {year} {2019}{\natexlab{c}})}\BibitemShut {NoStop}%
\bibitem [{\citenamefont {Mendelson}\ \emph {et~al.}(2020)\citenamefont
  {Mendelson}, \citenamefont {Chugh}, \citenamefont {Reimers}, \citenamefont
  {Cheng}, \citenamefont {Gottscholl}, \citenamefont {Long}, \citenamefont
  {Mellor}, \citenamefont {Zettl}, \citenamefont {Dyakonov}, \citenamefont
  {Beton}, \citenamefont {Novikov}, \citenamefont {Jagadish}, \citenamefont
  {Tan}, \citenamefont {Ford}, \citenamefont {Toth}, \citenamefont {Bradac},\
  and\ \citenamefont {Aharonovich}}]{arXiv:2003.00949}%
  \BibitemOpen
  \bibfield  {author} {\bibinfo {author} {\bibfnamefont {Noah}\ \bibnamefont
  {Mendelson}}, \bibinfo {author} {\bibfnamefont {Dipankar}\ \bibnamefont
  {Chugh}}, \bibinfo {author} {\bibfnamefont {Jeffrey~R.}\ \bibnamefont
  {Reimers}}, \bibinfo {author} {\bibfnamefont {Tin~S.}\ \bibnamefont {Cheng}},
  \bibinfo {author} {\bibfnamefont {Andreas}\ \bibnamefont {Gottscholl}},
  \bibinfo {author} {\bibfnamefont {Hu}~\bibnamefont {Long}}, \bibinfo {author}
  {\bibfnamefont {Christopher~J.}\ \bibnamefont {Mellor}}, \bibinfo {author}
  {\bibfnamefont {Alex}\ \bibnamefont {Zettl}}, \bibinfo {author}
  {\bibfnamefont {Vladimir}\ \bibnamefont {Dyakonov}}, \bibinfo {author}
  {\bibfnamefont {Peter~H.}\ \bibnamefont {Beton}}, \bibinfo {author}
  {\bibfnamefont {Sergei~V.}\ \bibnamefont {Novikov}}, \bibinfo {author}
  {\bibfnamefont {Chennupati}\ \bibnamefont {Jagadish}}, \bibinfo {author}
  {\bibfnamefont {Hark~Hoe}\ \bibnamefont {Tan}}, \bibinfo {author}
  {\bibfnamefont {Michael~J.}\ \bibnamefont {Ford}}, \bibinfo {author}
  {\bibfnamefont {Milos}\ \bibnamefont {Toth}}, \bibinfo {author}
  {\bibfnamefont {Carlo}\ \bibnamefont {Bradac}}, \ and\ \bibinfo {author}
  {\bibfnamefont {Igor}\ \bibnamefont {Aharonovich}},\ }\bibfield  {title}
  {\enquote {\bibinfo {title} {Identifying carbon as the source of visible
  single photon emission from hexagonal boron nitride},}\ }\href
  {https://arxiv.org/abs/2003.00949} {\bibfield  {journal} {\bibinfo  {journal}
  {arXiv:2003.00949}\ } (\bibinfo {year} {2020})}\BibitemShut {NoStop}%
\bibitem [{\citenamefont {Liu}\ \emph {et~al.}(2020)\citenamefont {Liu},
  \citenamefont {Wang}, \citenamefont {Li}, \citenamefont {Yu}, \citenamefont
  {Ke}, \citenamefont {Meng}, \citenamefont {Tang}, \citenamefont {Li},\ and\
  \citenamefont {Guo}}]{LIU2020114251}%
  \BibitemOpen
  \bibfield  {author} {\bibinfo {author} {\bibfnamefont {Wei}\ \bibnamefont
  {Liu}}, \bibinfo {author} {\bibfnamefont {Yi-Tao}\ \bibnamefont {Wang}},
  \bibinfo {author} {\bibfnamefont {Zhi-Peng}\ \bibnamefont {Li}}, \bibinfo
  {author} {\bibfnamefont {Shang}\ \bibnamefont {Yu}}, \bibinfo {author}
  {\bibfnamefont {Zhi-Jin}\ \bibnamefont {Ke}}, \bibinfo {author}
  {\bibfnamefont {Yu}~\bibnamefont {Meng}}, \bibinfo {author} {\bibfnamefont
  {Jian-Shun}\ \bibnamefont {Tang}}, \bibinfo {author} {\bibfnamefont
  {Chuan-Feng}\ \bibnamefont {Li}}, \ and\ \bibinfo {author} {\bibfnamefont
  {Guang-Can}\ \bibnamefont {Guo}},\ }\bibfield  {title} {\enquote {\bibinfo
  {title} {An ultrastable and robust single-photon emitter in hexagonal boron
  nitride},}\ }\href {\doibase https://doi.org/10.1016/j.physe.2020.114251}
  {\bibfield  {journal} {\bibinfo  {journal} {Physica E Low Dimens. Syst.
  Nanostruct.}\ }\textbf {\bibinfo {volume} {124}},\ \bibinfo {pages} {114251}
  (\bibinfo {year} {2020})}\BibitemShut {NoStop}%
\bibitem [{\citenamefont {Chejanovsky}\ \emph {et~al.}(2016)\citenamefont
  {Chejanovsky}, \citenamefont {Rezai}, \citenamefont {Paolucci}, \citenamefont
  {Kim}, \citenamefont {Rendler}, \citenamefont {Rouabeh}, \citenamefont
  {F{\'a}varo~de Oliveira}, \citenamefont {Herlinger}, \citenamefont
  {Denisenko}, \citenamefont {Yang}, \citenamefont {Gerhardt}, \citenamefont
  {Finkler}, \citenamefont {Smet},\ and\ \citenamefont
  {Wrachtrup}}]{10.1021/acs.nanolett.6b03268}%
  \BibitemOpen
  \bibfield  {author} {\bibinfo {author} {\bibfnamefont {Nathan}\ \bibnamefont
  {Chejanovsky}}, \bibinfo {author} {\bibfnamefont {Mohammad}\ \bibnamefont
  {Rezai}}, \bibinfo {author} {\bibfnamefont {Federico}\ \bibnamefont
  {Paolucci}}, \bibinfo {author} {\bibfnamefont {Youngwook}\ \bibnamefont
  {Kim}}, \bibinfo {author} {\bibfnamefont {Torsten}\ \bibnamefont {Rendler}},
  \bibinfo {author} {\bibfnamefont {Wafa}\ \bibnamefont {Rouabeh}}, \bibinfo
  {author} {\bibfnamefont {Felipe}\ \bibnamefont {F{\'a}varo~de Oliveira}},
  \bibinfo {author} {\bibfnamefont {Patrick}\ \bibnamefont {Herlinger}},
  \bibinfo {author} {\bibfnamefont {Andrej}\ \bibnamefont {Denisenko}},
  \bibinfo {author} {\bibfnamefont {Sen}\ \bibnamefont {Yang}}, \bibinfo
  {author} {\bibfnamefont {Ilja}\ \bibnamefont {Gerhardt}}, \bibinfo {author}
  {\bibfnamefont {Amit}\ \bibnamefont {Finkler}}, \bibinfo {author}
  {\bibfnamefont {Jurgen~H.}\ \bibnamefont {Smet}}, \ and\ \bibinfo {author}
  {\bibfnamefont {Jörg}\ \bibnamefont {Wrachtrup}},\ }\bibfield  {title}
  {\enquote {\bibinfo {title} {Structural attributes and photodynamics of
  visible spectrum quantum emitters in hexagonal boron nitride},}\ }\href
  {\doibase 10.1021/acs.nanolett.6b03268} {\bibfield  {journal} {\bibinfo
  {journal} {Nano Lett.}\ }\textbf {\bibinfo {volume} {16}},\ \bibinfo {pages}
  {7037--7045} (\bibinfo {year} {2016})}\BibitemShut {NoStop}%
\bibitem [{\citenamefont {Choi}\ \emph {et~al.}(2016)\citenamefont {Choi},
  \citenamefont {Tran}, \citenamefont {Elbadawi}, \citenamefont {Lobo},
  \citenamefont {Wang}, \citenamefont {Juodkazis}, \citenamefont {Seniutinas},
  \citenamefont {Toth},\ and\ \citenamefont
  {Aharonovich}}]{10.1021/acsami.6b09875}%
  \BibitemOpen
  \bibfield  {author} {\bibinfo {author} {\bibfnamefont {Sumin}\ \bibnamefont
  {Choi}}, \bibinfo {author} {\bibfnamefont {Toan~Trong}\ \bibnamefont {Tran}},
  \bibinfo {author} {\bibfnamefont {Christopher}\ \bibnamefont {Elbadawi}},
  \bibinfo {author} {\bibfnamefont {Charlene}\ \bibnamefont {Lobo}}, \bibinfo
  {author} {\bibfnamefont {Xuewen}\ \bibnamefont {Wang}}, \bibinfo {author}
  {\bibfnamefont {Saulius}\ \bibnamefont {Juodkazis}}, \bibinfo {author}
  {\bibfnamefont {Gediminas}\ \bibnamefont {Seniutinas}}, \bibinfo {author}
  {\bibfnamefont {Milos}\ \bibnamefont {Toth}}, \ and\ \bibinfo {author}
  {\bibfnamefont {Igor}\ \bibnamefont {Aharonovich}},\ }\bibfield  {title}
  {\enquote {\bibinfo {title} {Engineering and localization of quantum emitters
  in large hexagonal boron nitride layers},}\ }\href {\doibase
  10.1021/acsami.6b09875} {\bibfield  {journal} {\bibinfo  {journal} {ACS Appl.
  Mater. Interfaces}\ }\textbf {\bibinfo {volume} {8}},\ \bibinfo {pages}
  {29642--29648} (\bibinfo {year} {2016})}\BibitemShut {NoStop}%
\bibitem [{\citenamefont {Dietrich}\ \emph {et~al.}(2018)\citenamefont
  {Dietrich}, \citenamefont {B\"urk}, \citenamefont {Steiger}, \citenamefont
  {Antoniuk}, \citenamefont {Tran}, \citenamefont {Nguyen}, \citenamefont
  {Aharonovich}, \citenamefont {Jelezko},\ and\ \citenamefont
  {Kubanek}}]{PhysRevB.98.081414}%
  \BibitemOpen
  \bibfield  {author} {\bibinfo {author} {\bibfnamefont {A.}~\bibnamefont
  {Dietrich}}, \bibinfo {author} {\bibfnamefont {M.}~\bibnamefont {B\"urk}},
  \bibinfo {author} {\bibfnamefont {E.~S.}\ \bibnamefont {Steiger}}, \bibinfo
  {author} {\bibfnamefont {L.}~\bibnamefont {Antoniuk}}, \bibinfo {author}
  {\bibfnamefont {T.~T.}\ \bibnamefont {Tran}}, \bibinfo {author}
  {\bibfnamefont {M.}~\bibnamefont {Nguyen}}, \bibinfo {author} {\bibfnamefont
  {I.}~\bibnamefont {Aharonovich}}, \bibinfo {author} {\bibfnamefont
  {F.}~\bibnamefont {Jelezko}}, \ and\ \bibinfo {author} {\bibfnamefont
  {A.}~\bibnamefont {Kubanek}},\ }\bibfield  {title} {\enquote {\bibinfo
  {title} {{Observation of Fourier transform limited lines in hexagonal boron
  nitride}},}\ }\href {\doibase 10.1103/PhysRevB.98.081414} {\bibfield
  {journal} {\bibinfo  {journal} {Phys. Rev. B}\ }\textbf {\bibinfo {volume}
  {98}},\ \bibinfo {pages} {081414(R)} (\bibinfo {year} {2018})}\BibitemShut
  {NoStop}%
\bibitem [{\citenamefont {Noh}\ \emph {et~al.}(2018)\citenamefont {Noh},
  \citenamefont {Choi}, \citenamefont {Kim}, \citenamefont {Im}, \citenamefont
  {Kim}, \citenamefont {Seo},\ and\ \citenamefont
  {Lee}}]{10.1021/acs.nanolett.8b01030}%
  \BibitemOpen
  \bibfield  {author} {\bibinfo {author} {\bibfnamefont {Gichang}\ \bibnamefont
  {Noh}}, \bibinfo {author} {\bibfnamefont {Daebok}\ \bibnamefont {Choi}},
  \bibinfo {author} {\bibfnamefont {Jin-Hun}\ \bibnamefont {Kim}}, \bibinfo
  {author} {\bibfnamefont {Dong-Gil}\ \bibnamefont {Im}}, \bibinfo {author}
  {\bibfnamefont {Yoon-Ho}\ \bibnamefont {Kim}}, \bibinfo {author}
  {\bibfnamefont {Hosung}\ \bibnamefont {Seo}}, \ and\ \bibinfo {author}
  {\bibfnamefont {Jieun}\ \bibnamefont {Lee}},\ }\bibfield  {title} {\enquote
  {\bibinfo {title} {Stark tuning of single-photon emitters in hexagonal boron
  nitride},}\ }\href {\doibase 10.1021/acs.nanolett.8b01030} {\bibfield
  {journal} {\bibinfo  {journal} {Nano Lett.}\ }\textbf {\bibinfo {volume}
  {18}},\ \bibinfo {pages} {4710--4715} (\bibinfo {year} {2018})}\BibitemShut
  {NoStop}%
\bibitem [{\citenamefont {Mendelson}\ \emph {et~al.}(2019)\citenamefont
  {Mendelson}, \citenamefont {Xu}, \citenamefont {Tran}, \citenamefont
  {Kianinia}, \citenamefont {Scott}, \citenamefont {Bradac}, \citenamefont
  {Aharonovich},\ and\ \citenamefont {Toth}}]{10.1021/acsnano.8b08511}%
  \BibitemOpen
  \bibfield  {author} {\bibinfo {author} {\bibfnamefont {Noah}\ \bibnamefont
  {Mendelson}}, \bibinfo {author} {\bibfnamefont {Zai-Quan}\ \bibnamefont
  {Xu}}, \bibinfo {author} {\bibfnamefont {Toan~Trong}\ \bibnamefont {Tran}},
  \bibinfo {author} {\bibfnamefont {Mehran}\ \bibnamefont {Kianinia}}, \bibinfo
  {author} {\bibfnamefont {John}\ \bibnamefont {Scott}}, \bibinfo {author}
  {\bibfnamefont {Carlo}\ \bibnamefont {Bradac}}, \bibinfo {author}
  {\bibfnamefont {Igor}\ \bibnamefont {Aharonovich}}, \ and\ \bibinfo {author}
  {\bibfnamefont {Milos}\ \bibnamefont {Toth}},\ }\bibfield  {title} {\enquote
  {\bibinfo {title} {Engineering and tuning of quantum emitters in few-layer
  hexagonal boron nitride},}\ }\href {\doibase 10.1021/acsnano.8b08511}
  {\bibfield  {journal} {\bibinfo  {journal} {ACS Nano}\ }\textbf {\bibinfo
  {volume} {13}},\ \bibinfo {pages} {3132--3140} (\bibinfo {year}
  {2019})}\BibitemShut {NoStop}%
\bibitem [{\citenamefont {White}\ \emph {et~al.}(2020)\citenamefont {White},
  \citenamefont {Klauck}, \citenamefont {Tran}, \citenamefont {Schmitt},
  \citenamefont {Kianinia}, \citenamefont {Steinfurth}, \citenamefont
  {Heinrich}, \citenamefont {Toth}, \citenamefont {Szameit}, \citenamefont
  {Aharonovich},\ and\ \citenamefont {Solntsev}}]{arXiv:2001.10625}%
  \BibitemOpen
  \bibfield  {author} {\bibinfo {author} {\bibfnamefont {Simon J.~U.}\
  \bibnamefont {White}}, \bibinfo {author} {\bibfnamefont {Friederike}\
  \bibnamefont {Klauck}}, \bibinfo {author} {\bibfnamefont {Toan~Trong}\
  \bibnamefont {Tran}}, \bibinfo {author} {\bibfnamefont {Nora}\ \bibnamefont
  {Schmitt}}, \bibinfo {author} {\bibfnamefont {Mehran}\ \bibnamefont
  {Kianinia}}, \bibinfo {author} {\bibfnamefont {Andrea}\ \bibnamefont
  {Steinfurth}}, \bibinfo {author} {\bibfnamefont {Matthias}\ \bibnamefont
  {Heinrich}}, \bibinfo {author} {\bibfnamefont {Milos}\ \bibnamefont {Toth}},
  \bibinfo {author} {\bibfnamefont {Alexander}\ \bibnamefont {Szameit}},
  \bibinfo {author} {\bibfnamefont {Igor}\ \bibnamefont {Aharonovich}}, \ and\
  \bibinfo {author} {\bibfnamefont {Alexander}\ \bibnamefont {Solntsev}},\
  }\bibfield  {title} {\enquote {\bibinfo {title} {Quantum random number
  generation using a solid-state single-photon source},}\ }\href
  {https://arxiv.org/abs/2001.10625} {\bibfield  {journal} {\bibinfo  {journal}
  {arXiv:2001.10625}\ } (\bibinfo {year} {2020})}\BibitemShut {NoStop}%
\bibitem [{\citenamefont {Chen}\ \emph {et~al.}(2020)\citenamefont {Chen},
  \citenamefont {Tran}, \citenamefont {Duong}, \citenamefont {Li},
  \citenamefont {Toth}, \citenamefont {Bradac}, \citenamefont {Aharonovich},
  \citenamefont {Solntsev},\ and\ \citenamefont
  {Tran}}]{10.1021/acsami.0c05735}%
  \BibitemOpen
  \bibfield  {author} {\bibinfo {author} {\bibfnamefont {Yongliang}\
  \bibnamefont {Chen}}, \bibinfo {author} {\bibfnamefont {Thinh~Ngoc}\
  \bibnamefont {Tran}}, \bibinfo {author} {\bibfnamefont {Ngoc My~Hanh}\
  \bibnamefont {Duong}}, \bibinfo {author} {\bibfnamefont {Chi}\ \bibnamefont
  {Li}}, \bibinfo {author} {\bibfnamefont {Milos}\ \bibnamefont {Toth}},
  \bibinfo {author} {\bibfnamefont {Carlo}\ \bibnamefont {Bradac}}, \bibinfo
  {author} {\bibfnamefont {Igor}\ \bibnamefont {Aharonovich}}, \bibinfo
  {author} {\bibfnamefont {Alexander}\ \bibnamefont {Solntsev}}, \ and\
  \bibinfo {author} {\bibfnamefont {Toan~Trong}\ \bibnamefont {Tran}},\
  }\bibfield  {title} {\enquote {\bibinfo {title} {Optical thermometry with
  quantum emitters in hexagonal boron nitride},}\ }\href {\doibase
  10.1021/acsami.0c05735} {\bibfield  {journal} {\bibinfo  {journal} {ACS Appl.
  Mater. Interfaces}\ }\textbf {\bibinfo {volume} {12}},\ \bibinfo {pages}
  {25464--25470} (\bibinfo {year} {2020})}\BibitemShut {NoStop}%
\bibitem [{\citenamefont {Sawant}\ \emph {et~al.}(2014)\citenamefont {Sawant},
  \citenamefont {Samuel}, \citenamefont {Sinha}, \citenamefont {Sinha},\ and\
  \citenamefont {Sinha}}]{PhysRevLett.113.120406}%
  \BibitemOpen
  \bibfield  {author} {\bibinfo {author} {\bibfnamefont {Rahul}\ \bibnamefont
  {Sawant}}, \bibinfo {author} {\bibfnamefont {Joseph}\ \bibnamefont {Samuel}},
  \bibinfo {author} {\bibfnamefont {Aninda}\ \bibnamefont {Sinha}}, \bibinfo
  {author} {\bibfnamefont {Supurna}\ \bibnamefont {Sinha}}, \ and\ \bibinfo
  {author} {\bibfnamefont {Urbasi}\ \bibnamefont {Sinha}},\ }\bibfield  {title}
  {\enquote {\bibinfo {title} {Nonclassical paths in quantum interference
  experiments},}\ }\href {\doibase 10.1103/PhysRevLett.113.120406} {\bibfield
  {journal} {\bibinfo  {journal} {Phys. Rev. Lett.}\ }\textbf {\bibinfo
  {volume} {113}},\ \bibinfo {pages} {120406} (\bibinfo {year}
  {2014})}\BibitemShut {NoStop}%
\bibitem [{\citenamefont {De~Raedt}\ \emph {et~al.}(2012)\citenamefont
  {De~Raedt}, \citenamefont {Michielsen},\ and\ \citenamefont
  {Hess}}]{PhysRevA.85.012101}%
  \BibitemOpen
  \bibfield  {author} {\bibinfo {author} {\bibfnamefont {Hans}\ \bibnamefont
  {De~Raedt}}, \bibinfo {author} {\bibfnamefont {Kristel}\ \bibnamefont
  {Michielsen}}, \ and\ \bibinfo {author} {\bibfnamefont {Karl}\ \bibnamefont
  {Hess}},\ }\bibfield  {title} {\enquote {\bibinfo {title} {Analysis of
  multipath interference in three-slit experiments},}\ }\href {\doibase
  10.1103/PhysRevA.85.012101} {\bibfield  {journal} {\bibinfo  {journal} {Phys.
  Rev. A}\ }\textbf {\bibinfo {volume} {85}},\ \bibinfo {pages} {012101}
  (\bibinfo {year} {2012})}\BibitemShut {NoStop}%
\bibitem [{\citenamefont {Qian}\ \emph {et~al.}(2020)\citenamefont {Qian},
  \citenamefont {Konthasinghe}, \citenamefont {Manikandan}, \citenamefont
  {Spiecker}, \citenamefont {Vamivakas},\ and\ \citenamefont
  {Eberly}}]{PhysRevResearch.2.012016}%
  \BibitemOpen
  \bibfield  {author} {\bibinfo {author} {\bibfnamefont {X.-F.}\ \bibnamefont
  {Qian}}, \bibinfo {author} {\bibfnamefont {K.}~\bibnamefont {Konthasinghe}},
  \bibinfo {author} {\bibfnamefont {S.~K.}\ \bibnamefont {Manikandan}},
  \bibinfo {author} {\bibfnamefont {D.}~\bibnamefont {Spiecker}}, \bibinfo
  {author} {\bibfnamefont {A.~N.}\ \bibnamefont {Vamivakas}}, \ and\ \bibinfo
  {author} {\bibfnamefont {J.~H.}\ \bibnamefont {Eberly}},\ }\bibfield  {title}
  {\enquote {\bibinfo {title} {Turning off quantum duality},}\ }\href {\doibase
  10.1103/PhysRevResearch.2.012016} {\bibfield  {journal} {\bibinfo  {journal}
  {Phys. Rev. Research}\ }\textbf {\bibinfo {volume} {2}},\ \bibinfo {pages}
  {012016} (\bibinfo {year} {2020})}\BibitemShut {NoStop}%
\bibitem [{\citenamefont {Gisin}\ \emph {et~al.}(2002)\citenamefont {Gisin},
  \citenamefont {Ribordy}, \citenamefont {Tittel},\ and\ \citenamefont
  {Zbinden}}]{RevModPhys.74.145}%
  \BibitemOpen
  \bibfield  {author} {\bibinfo {author} {\bibfnamefont {Nicolas}\ \bibnamefont
  {Gisin}}, \bibinfo {author} {\bibfnamefont {Gr\'egoire}\ \bibnamefont
  {Ribordy}}, \bibinfo {author} {\bibfnamefont {Wolfgang}\ \bibnamefont
  {Tittel}}, \ and\ \bibinfo {author} {\bibfnamefont {Hugo}\ \bibnamefont
  {Zbinden}},\ }\bibfield  {title} {\enquote {\bibinfo {title} {Quantum
  cryptography},}\ }\href {\doibase 10.1103/RevModPhys.74.145} {\bibfield
  {journal} {\bibinfo  {journal} {Rev. Mod. Phys.}\ }\textbf {\bibinfo {volume}
  {74}},\ \bibinfo {pages} {145--195} (\bibinfo {year} {2002})}\BibitemShut
  {NoStop}%
\bibitem [{\citenamefont {Knopf}\ \emph {et~al.}(2019)\citenamefont {Knopf},
  \citenamefont {Lundt}, \citenamefont {Bucher}, \citenamefont {H\"{o}fling},
  \citenamefont {Tongay}, \citenamefont {Taniguchi}, \citenamefont {Watanabe},
  \citenamefont {Staude}, \citenamefont {Schulz}, \citenamefont {Schneider},\
  and\ \citenamefont {Eilenberger}}]{10.1364/OME.9.000598}%
  \BibitemOpen
  \bibfield  {author} {\bibinfo {author} {\bibfnamefont {Heiko}\ \bibnamefont
  {Knopf}}, \bibinfo {author} {\bibfnamefont {Nils}\ \bibnamefont {Lundt}},
  \bibinfo {author} {\bibfnamefont {Tobias}\ \bibnamefont {Bucher}}, \bibinfo
  {author} {\bibfnamefont {Sven}\ \bibnamefont {H\"{o}fling}}, \bibinfo
  {author} {\bibfnamefont {Sefaattin}\ \bibnamefont {Tongay}}, \bibinfo
  {author} {\bibfnamefont {Takashi}\ \bibnamefont {Taniguchi}}, \bibinfo
  {author} {\bibfnamefont {Kenji}\ \bibnamefont {Watanabe}}, \bibinfo {author}
  {\bibfnamefont {Isabelle}\ \bibnamefont {Staude}}, \bibinfo {author}
  {\bibfnamefont {Ulrike}\ \bibnamefont {Schulz}}, \bibinfo {author}
  {\bibfnamefont {Christian}\ \bibnamefont {Schneider}}, \ and\ \bibinfo
  {author} {\bibfnamefont {Falk}\ \bibnamefont {Eilenberger}},\ }\bibfield
  {title} {\enquote {\bibinfo {title} {{Integration of atomically thin layers
  of transition metal dichalcogenides into high-Q, monolithic Bragg-cavities:
  an experimental platform for the enhancement of the optical interaction in
  2D-materials}},}\ }\href {\doibase 10.1364/OME.9.000598} {\bibfield
  {journal} {\bibinfo  {journal} {Opt. Mater. Express}\ }\textbf {\bibinfo
  {volume} {9}},\ \bibinfo {pages} {598--610} (\bibinfo {year}
  {2019})}\BibitemShut {NoStop}%
\bibitem [{\citenamefont {\ifmmode~\dot{Z}\else \.{Z}\fi{}ukowski}\ \emph
  {et~al.}(1997)\citenamefont {\ifmmode~\dot{Z}\else \.{Z}\fi{}ukowski},
  \citenamefont {Zeilinger},\ and\ \citenamefont {Horne}}]{PhysRevA.55.2564}%
  \BibitemOpen
  \bibfield  {author} {\bibinfo {author} {\bibfnamefont {Marek}\ \bibnamefont
  {\ifmmode~\dot{Z}\else \.{Z}\fi{}ukowski}}, \bibinfo {author} {\bibfnamefont
  {Anton}\ \bibnamefont {Zeilinger}}, \ and\ \bibinfo {author} {\bibfnamefont
  {Michael~A.}\ \bibnamefont {Horne}},\ }\bibfield  {title} {\enquote {\bibinfo
  {title} {Realizable higher-dimensional two-particle entanglements via
  multiport beam splitters},}\ }\href {\doibase 10.1103/PhysRevA.55.2564}
  {\bibfield  {journal} {\bibinfo  {journal} {Phys. Rev. A}\ }\textbf {\bibinfo
  {volume} {55}},\ \bibinfo {pages} {2564--2579} (\bibinfo {year}
  {1997})}\BibitemShut {NoStop}%
\end{thebibliography}

\end{document}